\documentclass[final]{article}

\usepackage{arxiv}

\usepackage[utf8]{inputenc} 
\usepackage[T1]{fontenc}    
\usepackage{hyperref}       
\usepackage{url}            
\usepackage{booktabs}       
\usepackage{amsfonts}       
\usepackage{nicefrac}       
\usepackage{microtype}      
\usepackage{lipsum}
\usepackage{graphicx}

\usepackage{xcolor}
\usepackage{tikz}
\usepackage{amsmath}
\usepackage{amssymb}

\usepackage{lineno}
\usepackage{algorithm2e}
\usepackage{textcomp} 
\usepackage{subcaption}
\usepackage{adjustbox}
\usepackage{booktabs}

\modulolinenumbers[5]
\usetikzlibrary{calc}
\usetikzlibrary{intersections}
\usetikzlibrary{through}

\title{Segmentation Algorithms for Ground-Based Infrared Cloud Images}

\author{
 Guillermo Terr\'en-Serrano \\
  Department of Electrical and Computer Engineering \\
  The University of New Mexico \\
  Albuquerque, NM 87131, United States\\
  \texttt{guillermoterren@unm.edu} \\
 \And
  Manel Mart\'inez-Ram\'on \\
  Department of Electrical and Computer Engineering \\
  The University of New Mexico \\
  Albuquerque, NM 87131, United States\\
  \texttt{manel@unm.edu} \\
}

\begin{document}

\maketitle

\begin{abstract}
    The increasing number of Photovoltaic (PV) systems connected to the power grid are vulnerable to the projection of shadows from moving clouds. Global Solar Irradiance (GSI) forecasting allows smart grids to optimize the energy dispatch, preventing energy shortages caused by occlusion of the sun. This investigation compares the performances of machine learning algorithms (not requiring labelled images for training) for real-time segmentation of clouds in images acquired using a ground-based infrared sky imager. Real-time segmentation is utilized to extract cloud features using only the pixels in which clouds are detected.
\end{abstract}


\section{Introduction}

Moving clouds produce reductions in the generation of energy from PV systems which are out of the grid operators acceptable range \cite{LAPPALAINEN2017}. Considering these circumstances, the implementation of intra-hour solar forecasting is necessary to attenuate the effects of energy reductions caused by clouds. 

Intra-hour solar forecasting (i.e. 15 minutes in advance) is used by grid operators for scheduling transmission services, so it is possible to control voltage fluctuations due to passing clouds \cite{URQUHART2013}. In this context, forecasting algorithms that do not include information extracted from sky images are not effective for intra-hour solar forecasting \cite{GARCIA2018}. The inclusion of cloudiness information from satellite images in a forecasting algorithm is useful when the horizon is between 15 minutes to an hour \cite{SOBRI2018}. For these reasons, ground-based sky imagers are the most suitable method for intra-hour solar forecasting \cite{KONG2020}.

When using visible light sky imagers, the pixels in the circumsolar area of the image are saturated. This is particularly problematic for intra-hour solar forecasting algorithms, because relevant forecasting information is suppressed in the pixels that are saturated. Blocking the Sun in the sky-images is a solution, but information is removed in the part of the image that is blocked by the shade structure. The same problem occurs when a total sky imager is used. The saturated area in the images is reduced when using a thermal sky imager \cite{MAMMOLI2019}. Thermal imaging systems allow the derivation of the temperature and altitude of clouds \cite{Escrig2013}. In previous investigations, ground-based radiometric infrared sky imagers \cite{SHAW2013} have been utilized to analyze the dynamics of clouds \cite{THURAIRAJAH2007} and to establish visual links for Earth-space communications \cite{NUGENT2009}.

Cloud segmentation algorithms are necessary to reduce the noise in a solar forecasting algorithm \cite{BERNECKER2014}. Previous research regarding cloud segmentation has shown that the accuracy of the segmentation models increases when information extracted from neighboring pixels is included \cite{Shi2017}. Operations using the dense Gram matrix in kernel learning methods \cite{shawe2004} are a problem for real-time cloud segmentation \cite{Taravat2015}. While the super-pixel approach reduces the computation time, it produces a coarse segmentation \cite{Deng2019}. Convolutional neural networks may be used in cloud segmentation, but require a large amount of training samples to avoid overfitting \cite{Dronner2018}. Clustering methods such as DBSCAN \cite{Ester1996} and HDBSCAN \cite{mcinnes2017} are the state of the art for identifying classes in large datasets. However, we have prior knowledge of the number of clusters in this application, and thus the additional computation burden is unnecessary \cite{HAR2005}.

Deep learning is prone to overfitting when the number of samples is small \cite{Brigato2021}. In contrast, learning methods based on Bayes' theorem (i.e. Bayesian inference) regularize the parameters in the learning process. For this reason, the machine learning algorithms proposed in this investigation are the Gaussian Mixture Model (GMM), k-means and Markov Random Fields (MRF). 

GMM is a clustering algorithm that infers the distribution of the clusters using a dense covariance matrix. The training and testing time of a GMM may be improved by reducing the number of parameters in the covariance matrix that need to be inferred. In the case of k-means clustering, the covariance matrix is defined as an identify matrix \cite{MURPHY2012}. MRF models are computationally expensive but suitable for segmentation problems, because information from the classification of neighboring pixels is included in the prior. The Iterated Conditional Modes (ICM) algorithm allows for training of MRF models in an unsupervised manner. With the aim of reducing the classification time, the Simulated Annealing (SA) algorithm is implemented to perform an efficient optimization of the ICM-MRF.

\section{Dataset}

This data was acquired by a sky imager mounted on a solar tracker which maintains the Sun in the center of the images. The sky imager is equipped with a Lepton 2.5 radiometric infrared camera that measures temperature in centi-kelvin units. The resolution of the camera is $80 \times 60$ pixels and the diagonal FOV is $60^\circ$. The thermal sky imager is located at the ECE department in the UNM Central Campus. The weather features are measured by a weather station located at the UNM Hospital.

\subsection{Infrared Images}

A pixel of the camera frame is defined by its pair of Euclidean coordinates $i,j$. The temperatures in an image are defined as $\mathbf{T} = \{ T_{i,j} \in \mathbb{R}^+ \mid \forall i = 1, \ldots, M, \ \forall j = 1, \ldots, N \}$. The height of the pixels $\mathbf{H} = \{ H_{i,j} \in \mathbb{R}^+ \mid \forall i = 1, \ldots, M, \ \forall j = 1, \ldots, N \}$ is computed using the Moist Adiabatic Lapse Rate (MALR) \cite{HOLTON2013}. The MALR approximates the decrease of the temperature in the troposphere as a function of the air temperature, dew point and relative humidity in the ground. The images are preprocessed to remove stationary artifacts such as water stains \cite{TERREN2020a}. The temperature of the pixels after applying the preprocessing algorithm are $\mathbf{T}^\prime = \{ T^\prime_{i,j} \in \mathbb{R}^+ \mid \ \forall i = 1, \ldots, M, \ \forall j = 1, \ldots, N \}$, and the heights are $\mathbf{H}^{\prime} = \{ H_{i,j}^\prime \in \mathbb{R}^+ \mid \forall i = 1, \ldots, M, \ \forall j = 1, \ldots, N \}$. After preprocessing the images to remove stationary artifacts, they are preprocessed again to remove the effect of the Sun and the atmospheric background radiation. The obtained values of the pixels are the difference of temperature with respect to the Tropopause, defined as $\mathbf{\Delta T} = \{ \Delta T_{i,j} \in \mathbb{R} \mid \forall i = 1, \ldots, M, \ \forall j = 1, \ldots, N \}$. The temperature differences are multiplied by the average atmospheric background temperature to compute the heights: $\mathbf{H}^{\prime \prime} = \{ H_{i,j}^{\prime \prime} \in \mathbb{R}^+ \mid \forall i = 1, \ldots, M, \ \forall j = 1, \ldots, N \}$. After applying both preprocessing algorithms, the temperature differences are normalized to an 8 bit image $\bar{\mathbf{T}} = \{ \bar{T}_{i,j} \in \mathbb{N}^{2^{8}} \mid \forall i = 1, \ldots, M, \ \forall j = 1, \ldots, N \}$, calculating the maximum feasible temperature of a cloud within the Tropopause \cite{TERREN2020a}, and assuming a temperature decrease of $9.8^\circ/$km \cite{Hummel1981}. The velocity vectors were computed applying the Lucas-Kanade algorithm to two consecutive normalized images \cite{Lucas1981}.

\subsection{Feature Vectors}

The four different combinations of features extracted from the images are used as input vectors in the segmentation models. The first, second, third and fourth vectors are: $\mathbf{x}^{1}_{i,j} = \{ T_{i,j}, \ H_{i,j} \}$, $\mathbf{x}^{2}_{i,j} = \{ T_{i,j}^\prime, \ H_{i,j}^\prime \}$, $\mathbf{x}^{3}_{i,j} = \{ \Delta T_{i,j}, \ H_{i,j}^{\prime \prime} \}$ and $\mathbf{x}^{4}_{i,j} = \{ \mathrm{mag} (\mathbf{v}_{i,j}), \ \bar{T}_{i,j}, \ \Delta T_{i,j} \}$ respectively. Other combinations of feature vectors were explored and were found to under-perform in cloud classification. Other combinations of feature vectors were explored and were found to under-perform in cloud classification. 

Features extracted from neighboring pixels are included in the feature vectors of pixel $i,j$:
\begin{itemize}
    \item Single pixel: $\{ \mathbf{x}_{i,j} \}, \quad \forall i,j = i_1,j_1, \ldots, i_{M}, j_{N}$

    \item $1^{st}$ order: $\{\mathbf{x}_{i-1,j}, \ \mathbf{x}_{i,j - 1}, \ \mathbf{x}_{i,j + 1}, \ \mathbf{x}_{i+1,j} \}$.

    \item $2^{nd}$ order: $\{\mathbf{x}_{i-1,j}, \ \mathbf{x}_{i,j - 1}, \ \mathbf{x}_{i,j + 1}, \ \mathbf{x}_{i + 1,j}, \ \mathbf{x}_{i - 1,j -1}, \dots$ $\dots, \ \mathbf{x}_{i - 1,j + 1}, \ \mathbf{x}_{i + 1,j + 1}, \ \mathbf{x}_{i + 1,j - 1} \}$.
\end{itemize}

When single pixels are used, the vectors do not contain features from other pixels. The \emph{$1^{th}$ order neighborhood} contains features of the 4 closest pixels. The \emph{$2^{nd}$ order neighborhood} contains features of the 8 closest pixels.

\section{Methods}

The methods described below can be classified as generative when they have the capacity of generating new samples from a likelihood model, this is, when the model implements a density approximation of the form $p ( {\bf x} | \mathcal{C}_k )$ where $\mathcal{C}_k $ is the segmentation label of the pixel. The dataset is defined as $\mathcal{D} = \{ \mathbf{X}, \ \mathbf{y} \}$, where $\mathbf{x}_i \in \mathbb{R}^{d}$ and $y_i \in \{ -1, \ 1\}$ which are labels for a clear or cloudy pixel, respectively.

\subsection{Gaussian Mixture Model}

Feature distributions can be approximate by a mixture of multivariate normal distributions $\mathbf{x}_i \sim \mathcal{N} ( \boldsymbol{\mu}_k, \boldsymbol{\Sigma}_k )$. Under the hypothesis that a sample $\mathbf{x}_i$ belongs to class $\mathcal{C}_k$, its class conditional likelihood is 
\begin{equation}
    \label{eq:EM_likelihood}
    f \left(\mathbf{x} ; \boldsymbol{\mu}_k, \boldsymbol{\Sigma}_k \right) = \frac{1}{\sqrt{ \left(2 \pi\right)^d \left| \boldsymbol{\Sigma}_k \right| }} \mathrm{e}^{ \left\{ - \frac{1}{2} ( \mathbf{x} - \boldsymbol{\mu}_k )^\top \boldsymbol{\Sigma}_k^{-1} ( \mathbf{x} - \boldsymbol{\mu}_k ) \right\}},
\end{equation}
where $\boldsymbol{\Sigma}_k$ is regularized to avoid an ill-conditioned covariance matrix such as $\boldsymbol{\Sigma}_k \triangleq \boldsymbol{\Sigma}_k + \varepsilon \mathbf{I}_{d \times d}$, $\varepsilon$ is the regularization hyperparameter. The number of distributions $k$ (i.e. clusters) is equivalent to the number of classes $\mathcal{C}_k$, henceforth $k = 2$ in our application (i.e. clear or cloudy pixel).

The expected complete data log-likelihood is defined as \cite{MURPHY2012},
\begin{equation}
    \mathcal{Q} ( \boldsymbol{\theta}^{t}, \boldsymbol{\theta}^{t-1}) = \sum_{i = 1}^N \sum_{i = k}^K \gamma_{i,k} \log \pi_k + \sum_{i = 1}^N \sum_{i = k}^K \gamma_{i,k}  \log p ( \mathbf{x}_i \mid \boldsymbol{\theta}^{t} )
\end{equation}
where $\gamma_{i,k} \triangleq p ( y_i = k \mid \mathbf{x}_i, \boldsymbol{\theta}^{t - 1} )$ is the responsibility of the cluster $k$ in the sample $i$.

The parameters in the clustering of multivariate normal distributions can be directly computed applying the Expectation Maximization (EM) algorithm. In the E stage of the algorithm a prior is established and then, by using the likelihood function \eqref{eq:EM_likelihood}, a posterior $\gamma_{i,k} = p ( \mathcal{C}_k|{\bf x}_i )$ can be assigned to each sample. In the M stage, the mean and variance of each cluster that maximize the log likelihood are computed as
\begin{equation}
    \begin{split}
        \label{eq:maximization_step}
        \boldsymbol{\mu}_k &= \frac{\sum_{i = 1}^N \gamma_{i,k} \mathbf{x}_i}{\gamma_k}, \\
        \boldsymbol{\Sigma}_k & = \frac{\sum_{i = 1}^N \gamma_{i,k} \mathbf{x}_i \mathbf{x}_i^\top}{\gamma_k} - \boldsymbol{\mu}_k \boldsymbol{\mu}_k^\top.   
    \end{split}
\end{equation}
The priors are updated as well using the posterior probabilities that are
\begin{equation}
    \pi_k = p(\mathcal{C}_k)=\frac{1}{N}\sum_{i = 1}^N \gamma_{i,k},
\end{equation}
where $N$ is the number of available samples. A class is assigned to each sample by Maximum A Posteriori (MAP) criteria $ \hat{y}_i = \underset{k}{\operatorname{argmax}} \ p \left( \mathcal{C}_k \mid \mathbf{x}_i, \boldsymbol{\mu}_k, \boldsymbol{\Sigma}_k \right)$
    
The theory behind mixture models, as well as the EM algorithm, is fully developed in \cite{MURPHY2012}. 

\subsection{k-means}

The k-means algorithm can be seen as a particularization of the algorithm above, where the posteriors $\gamma_{i,k}$  are  approximated by 1 if  distance $\|{\bf x}_i-{\boldsymbol \mu}_k\|<\|{\bf x}_i-{\boldsymbol \mu}_{k'}\|$, $k\neq k'$, and zero otherwise. The mean is computed as in Eq. \eqref{eq:maximization_step}, and the covariance is approximated by an identity matrix.

\subsection{Markov Random Fields}

The energy function of a MRF is composed of two functions \cite{STAN2001}. The function $\varphi$ that is the joint distribution of a class, and the function $\psi$ that is the potential energy of the system's configuration (a term from statistical mechanics),
\begin{equation}
    \label{eq:energy_function1}
    \mathcal{E} \left( y_i, \mathbf{x}_i \right) = \sum_{i} \varphi\left(\mathbf{x}_i, y_i \right) + \sum_{i,j} \psi \left( y_i, y_j \right),
\end{equation}
where $\mathbf{x}_i$ is the feature vector of sample $i$ and $y_i$ is its class. In the graph $G$, a sample $i$ has a set of neighboring pixels, and each neighboring sample $j$ has class $y_j$.

Sample $\mathbf{x}_i$ is classified using the Bayes' theorem as
\begin{equation}
    \label{eq:Bayes_theorem}
        p \left( y_{i} = \mathcal{C}_k \mid \mathbf{x}_{i}, \boldsymbol{\theta}_k \right)  
        \propto p \left( \mathbf{x}_{i} \mid y_{i} = \mathcal{C}_k, \boldsymbol{\theta}_k  \right) p \left( y_{i} = \mathcal{C}_k \right).
\end{equation}
where the corresponding likelihood is approximated by a normal distribution $\mathbf{x}_i \sim \mathcal{N} (\boldsymbol{\mu}_k, \boldsymbol{\Sigma}_k)$ of class $\mathcal{C}_k$, 
and $\boldsymbol{\theta}_k = \{\boldsymbol{\mu}_k, \boldsymbol{\Sigma}_k \}$ are the parameter set of the feature distribution in class $\mathcal{C}_k$.
The log-likelihood of class $\mathcal{C}_k$ is defined as $\varphi\left(\mathbf{x}_i, y_i \right) \triangleq \log p\left( \mathbf{x}_{i} | y_{i} = \mathcal{C}_k, \boldsymbol{\theta}_k \right)$ in the energy function \eqref{eq:energy_function1}. The prior can be expressed as,
\begin{equation}
    \label{eq:prior}
    p \left( y_{i} \right) = \frac{1}{Z} \exp \left( - \psi \left( y_{i}  \right) \right), 
\end{equation}
where $Z$ is the partition function for normalization. By applying the Hammersley–Clifford theorem \cite{HAMMERSLEY1971}, the potential function $\psi(y_i)$ in the exponential form can be factorized in cliques of a graph $G$. A clique is defined as a set of nodes that are all neighbors of each other \cite{MURPHY2012}. In this way, the potential function can be independently evaluated for each clique in the factorized graph,
\begin{equation}
    \label{eq:potential_function}
    \psi ( y_i ) = \sum_{\ell = 1}^L  \sum_{i,j \in \Omega_\ell} y_i \beta y_j ,
\end{equation}
where the set of maximal cliques in the graph is defined as $\Omega = \Omega_1 \cup \Omega_2 \cup \ldots \cup \Omega_L$, $\ell$ represents the order of the neighboring pixels to sample $i$ in the graph network $G$, and $\Omega_L$ is the maximal clique as it cannot be made any larger without losing the clique property \cite{MURPHY2012}. The cliques considered in our problem are $\Omega_1$ and $\Omega_2$, which  represent the $1^{st}$ and $2^{nd}$ order neighborhood cliques respectively. Hyperparameter $\beta$ needs to be cross-validated. 

By applying expression \eqref{eq:prior} in the logarithm of \eqref{eq:Bayes_theorem}, the energy function for a pixel $i$ of class $y_i$ and features ${\bf x}_i$ is
\begin{equation} 
    \label{eq:energy_function}
    \begin{split}
        \mathcal{E} &\left( y_{i} = \mathcal{C}_k \mid \mathbf{x}_{i}, \boldsymbol{\mu}_k, \boldsymbol{\Sigma}_k \right) = \- \frac{1}{2}\log \left| \boldsymbol{\Sigma}_k \right| \\
        & - \frac{1}{2}\left( \mathbf{x}_{i} - \boldsymbol{\mu}_k \right)^\top \boldsymbol{\Sigma}_k^{-1} \left( \mathbf{x}_{i} - \boldsymbol{\mu}_k \right) + \psi(y_i).
    \end{split}
\end{equation}
plus constant terms. Similarly to Eq. \eqref{eq:EM_likelihood}, the covariance matrix $\boldsymbol{\Sigma}_k$ in a MRF is also regularized as $\boldsymbol{\Sigma}_k \triangleq \boldsymbol{\Sigma}_k + \varepsilon \mathbf{I}_{d \times d}$. Finally, probability \eqref{eq:Bayes_theorem} can be written as
\begin{equation}
    p \left( y_i = \mathcal{C}_k \mid \mathbf{x}_i, \boldsymbol{\theta}_k \right) = \frac{\exp \ \mathcal{E} \left( y_i = \mathcal{C}_k \mid \mathbf{x}_i, \boldsymbol{\theta}_k \right)}{\sum_{k = 1}^K \exp \ \mathcal{E} \left( y_i = \mathcal{C}_k \mid \mathbf{x}_i \boldsymbol{\theta}_k \right) }.
\end{equation}
A class $\mathcal{C}_k$ is assigned to the sample $\mathbf{x}_i$ by the MAP criterion.

\subsubsection{Iterated Conditional Modes}

Parameters $\boldsymbol{\theta}_k$ in a MRF can be inferred with the ICM algorithm \cite{BESAG1986}. The algorithm initially assigns a class to each pixel from a uniform distribution. The samples with label $\mathcal{C}_k$ are defined within the set $S_k^{(0)}$. At iteration $t+1$ the mean and covariance of a class are computed as
\begin{equation}
    \begin{split}
        \boldsymbol{\mu}_k^{t+1} &= \frac{1}{\left|\mathcal{S}^{t}_k \right|} \sum_{\mathbf{x}_{i,j,z} \in S_k^{t}} \mathbf{x}_{i,j,z}, \\
        \boldsymbol{\Sigma}_k^{t+1} &= \frac{1}{\left| \mathcal{S}^{t}_k \right| - 1} \sum_{\mathbf{x}_{i,j} \in S_k^{t}} \left(\mathbf{x}_{i,j,z} - \boldsymbol{\mu}^{t+1}_k \right)^\top \left( \mathbf{x}_{i,j,z} - \boldsymbol{\mu}^{t+1}_k \right).
    \end{split}
\end{equation}

A class is reassigned to each pixel according to the parameters computed at iteration $t + 1$ with the MAP criterion
\begin{equation}
    y_{i,j}^{t+1} = \underset{k}{\operatorname{argmax}} \ \mathcal{E} \left( y_{i,j}^{t} \mid \mathbf{x}_{i,j}, \ \boldsymbol{\mu}_k^{t+1}, \boldsymbol{\Sigma}_k^{t+1} \right).
\end{equation}
When the total energy stops increasing, so that $\sum_{i,j}\mathcal{E} ( y_{i,j}^{t+1} \mid \mathbf{x}_{i,j}, \boldsymbol{\theta}_k^{t+1} ) \leq \sum_{i,j} \mathcal{E} ( y_{i,j}^{t} \mid \mathbf{x}_{i,j}, \boldsymbol{\theta}_k^{t} )$, the algorithm has converged to a stable configuration and the optimal set of parameters $\boldsymbol{\theta}_k$ have been found. The distribution of class $\mathcal{C}_k$ is defined as $\mathcal{N} ( \boldsymbol{\mu}_k^{t}, \boldsymbol{\Sigma}_k^{t} )$.

\subsubsection{Simulated Annealing}

The standard optimization goes through all the pixels calculating their potential and classifying them in each iteration of the algorithm. The computational cost of this method is high, but we can assume that it is not necessary to evaluate the pixels whose state has high energy, because their classification will not change. The computation cost can be reduced by sampling the pixels that are likely to be misclassified, and applying the optimization procedure only to them.

We propose to optimize the configuration of the pixels in an IR image applying the Simulated Annealing algorithm (SA) \cite{KIRK1983} to the MRF models \cite{KATO2001}. SA algorithm is applied on the implementation, after the inference of the class distributions. 

The class distributions $ \mathcal{N} ( \boldsymbol{\mu}_k, \mathbf{\Sigma}_k )$ were previously inferred applying a supervised or unsupervised learning algorithm. The optimization is initialized to maximum likelihood classification of the pixels $y_{i,j}^{\left( 0 \right)} = \underset{k}{\operatorname{argmax}} \ p \left( y_{i,j} = \mathcal{C}_k \mid \mathbf{x}_{i,j}, \boldsymbol{\theta}_k \right)$.

The likelihood a pixel to belong a class $\mathcal{C}_k$ is only evaluated at the initialization of the algorithm.

The objective is to evaluate the potential function of the samples that have low energy. For that, a sample $\mathbf{x}_{i,j}$ with label $y_{i,j} = \mathcal{C}_k$ is randomly selected and its classification is changed in each iteration $t$, so that $\bar{y}^{t}_{i,j} = - y_{i,j}^{t}$. The probability of selecting a sample $\mathbf{x}_{i,j}$ is weighted by their energy. The weights of the samples in an image are defined as,
\begin{equation}
    w_{i,j} = \frac{\mathcal{E} \left( \bar{y}^{t}_{i,j} \mid \mathbf{x}^{t}_{i,j}, \boldsymbol{\theta}_k \right) - \max_k \mathcal{E} \left( \bar{y}^{t}_{i,j} \mid \mathbf{x}^{t}_{i,j}, \boldsymbol{\theta}_k \right)}{\sum_{i,j} \left[ \mathcal{E} \left( \bar{y}^{t}_{i,j} \mid \mathbf{x}^{t}_{i,j}, \boldsymbol{\theta}_k \right) - \max_k \mathcal{E} \left( \bar{y}^{t}_{i,j} \mid \mathbf{x}^{t}_{i,j}, \boldsymbol{\theta}_k \right) \right]},
\end{equation}
and the cumulative distribution of the weights is computed such as $\bar{w}_{n,m} = \{ \{ \sum_{i = 1}^n \sum_{j = 1}^m w_{i,j} \}^{N}_{n = 1} \}^M_{m = 1}$. Then, a sample is drawn from a uniform distribution $\hat{w} \sim \mathcal{U} ( 0, 1 )$. The sample whose weight has the minimum distance to the drawn sample, is selected $i,j = \operatorname{argmin} | \bar{w}_{i,j} - \hat{w} |$.

The algorithm follows with Metropolis step which is computed with the energy of the changed sample  $\bar{y}_{i,j}$ and the energy of the original label $y_{i,j}$, $\Delta E = \mathcal{E} ( y_{i,j}^{t} \mid \mathbf{x}_{i,j}, \boldsymbol{\theta}_k  ) - \mathcal{E} ( \bar{y}_{i,j}^{t} \mid \mathbf{x}_{i,j}, \boldsymbol{\theta}_k )$.

The new label is directly accepted $\bar{y}_{i,j}^{t}$ iff $\Delta E < 0$. Otherwise, it will be accepted with probability $\rho = \exp (- \Delta E / T^{t} )$ in an analogous way to thermodynamics with the Gibbs distribution,
\begin{equation}
    \label{eq:acceptance_function}
    y_{i,j}^{t+1} = \left\{\begin{array}{ll}{ \bar{y}^{t}_{i,j} } & {\text { if } \Delta E \leq 0} \\ \bar{y}^{t}_{i,j} & {\text { if } \Delta E > 0 \text { and } \rho > u} \\ y_{i,j}^{t} & {\text { Otherwise}}\end{array}\right.
\end{equation}
the acceptance probability is drawn from a uniform distribution $u \sim \mathcal{U} \left(0, 1\right)$.

We propose to linearly cool down the acceptance rate through the temperature hyperparameter, so that $T^{t+1} = \alpha T^{t}$. The optimal hyperparameter $\alpha$ is a trade off between accuracy and speed.

\section{J-Statistic}

Younde's j-statistic \cite{YOUDEN1950}, is a test that evaluates the performances of a binary classification, it is defined as $ J = sensitivity + specificity - 1$.

To compensate possible class imbalance, we define a prior of the classification function $\lambda$ that is validated in each model, 
\begin{equation}
    \label{eq:virtual_prior}
        p \left( \mathcal{D} \mid \mathcal{C}_k \right) = \frac{p \left( \mathcal{C}_k \mid \mathcal{D} \right) p \left( \mathcal{C}_k  \right)}{p \left( \mathcal{D} \right)} 
        \propto p \left( \mathcal{C}_k \mid \mathcal{D}  \right) \lambda.
\end{equation}
The classification probabilities are defined as $p \left( \mathcal{D} \mid \mathcal{C}_1 \right) = p \left( \mathcal{C}_1 \mid \mathcal{D}  \right) \lambda $, and $ p \left( \mathcal{D} \mid \mathcal{C}_2 \right) = 1 - p \left( \mathcal{D} \mid \mathcal{C}_1 \right)$. The j-statistic score is maximized finding the optimal $\lambda$ threshold for binary classification. For that, the j-statistic is applied to the conventional Receiver Operating Characteristic (ROC) analysis \cite{FAWCETT2006}, and it is computed at each point of the ROC. We propose to use the maximum value of j-statistic in the ROC curve as the optimal point, so that $\hat{y}_* = \underset{k}{\operatorname{argmax}} \ p \left( \mathcal{C}_k \mid \mathbf{x}_*, \mathcal{D} \right) \lambda$.

\section{Results and Discussion}

The dataset used in this investigation was constructed using images in a publicly accessible database. The database of infrared sky images is available in a Dryad repository and described in \cite{TERREN2020d}. The software developed to performed the experiments is also available in a GitHub repository\footnote{\url{https://github.com/gterren/cloud_segmentation.git}}.

\begin{table}[!htbp]
    \centering
    \small
    \setlength{\tabcolsep}{12.pt} 
    \renewcommand{\arraystretch}{.25} 
    \caption{This table shows the j-statistics and average testing time archived by the different models. The j-statistics and average testing time are organized by neighborhood. ICM-MRF and SA-ICM-MRF models have two groups of vectors: those with a potential function of $1^{st}$ order $\Omega_1 (\cdot)$ or $2^{nd}$ order cliques $\Omega_2 (\cdot)$.}
    \begin{tabular}{lcccccc}
        \toprule
        {} & \multicolumn{3}{c}{J-statistic [\%]} & \multicolumn{3}{c}{Time [ms]} \\
        Feature Vector & Single & $1^{st}$ Order & $2^{nd}$ Order & Single & $1^{st}$ Order & $2^{nd}$ Order \\
        \midrule
        \multicolumn{7}{c}{k-means} \\
        \midrule
        $\mathbf{x}^1_{i,j}$ & 47.91 & 43.49 & 43.05 & 0.88 & 1.01 & 1.05 \\
        $\mathbf{x}^2_{i,j}$ & 48.32 & 46.67 & 42.87 & 0.85 & 1.00 & 1.03 \\
        $\mathbf{x}^3_{i,j}$ & 73.68 & \textbf{80.73} & 79.21 & 0.87 & \textbf{0.92} & 1.04 \\
        $\mathbf{x}^4_{i,j}$ & 78.52 & 78.19 & 78.10 & 0.83 & 1.04 & 1.15 \\
        \midrule
        \multicolumn{7}{c}{GMM} \\
        \midrule 
        $\mathbf{x}^1_{i,j}$ & 46.89 & 47.75 & 44.19 & 1.78 & 3.84 & 2.93 \\
        $\mathbf{x}^2_{i,j}$ & 48.46 & 53.70 & 53.30 & 1.77 & 2.24 & 4.25 \\
        $\mathbf{x}^3_{i,j}$ & 75.58 & 86.58 & 83.38 & 1.72 & 2.38 & 3.06 \\
        $\mathbf{x}^4_{i,j}$ & \textbf{89.39} & 88.55 & 83.93 & \textbf{4.33} & 2.75 & 3.88 \\
        \midrule
        \multicolumn{7}{c}{ICM-MRF} \\
        \midrule 
        $\Omega_1(\mathbf{x}^1_{i,j})$ & 20.00 &  31.67 & 20.00 & 141.44 & 135.23 & 422.62 \\
        $\Omega_1(\mathbf{x}^2_{i,j})$ & 23.36 & 32.47 & 20.00 & 143.62 & 242.50 & 374.71 \\
        $\Omega_1(\mathbf{x}^3_{i,j})$ & 71.80 & \textbf{92.55} & 90.61 & 268.60 & \textbf{640.81} & 291.18 \\
        $\Omega_1(\mathbf{x}^4_{i,j})$ & 57.02 & 72.25 & 71.71 & 271.48 & 661.79 & 218.39 \\
        $\Omega_2(\mathbf{x}^1_{i,j})$ & 22.14 & 31.54 & 20.00 & 250.78 & 242.01 & 743.55 \\
        $\Omega_2(\mathbf{x}^2_{i,j})$ & 20.00 & 32.04 & 19.05 & 397.42 & 258.82 & 634.30 \\
        $\Omega_2(\mathbf{x}^3_{i,j})$ & 46.15 & 74.37 & 70.12 & 638.44 & 2857.08 & 1262.52 \\
        $\Omega_2(\mathbf{x}^4_{i,j})$ & 61.37 & 70.04 & 90.94 & 336.07 & 2094.09 & 1007.52 \\
        \midrule 
        \multicolumn{7}{c}{SA-ICM-MRF} \\
        \midrule 
        $\Omega_1(\mathbf{x}^1_{i,j})$ & 20.00 & 20.66 & 20.81 & 128.26 & 129.98 & 131.74 \\
        $\Omega_1(\mathbf{x}^2_{i,j})$ & 20.00 & 20.66 & 20.73 & 127.25 & 129.72 & 132.93 \\
        $\Omega_1(\mathbf{x}^3_{i,j})$ & 67.70 & 90.01 & 68.83 & 129.52 & 130.63 & 131.93 \\
        $\Omega_1(\mathbf{x}^4_{i,j})$ & 82.98 & 70.17 & 68.05 & 132.04 & 131.88 & 133.45 \\
        $\Omega_2(\mathbf{x}^1_{i,j})$ & 20.07 & 20.66 & 20.53 & 136.63 & 138.73 & 138.74 \\
        $\Omega_2(\mathbf{x}^2_{i,j})$ & 20.10 & 20.67 & 20.73 & 135.30 & 144.00 & 139.47 \\
        $\Omega_2(\mathbf{x}^3_{i,j})$ & 52.67 & \textbf{90.06} & 68.80 & 135.63 & \textbf{136.53} & 139.32 \\
        $\Omega_2(\mathbf{x}^4_{i,j})$ & 68.92 & 73.70 & 72.57 & 136.86 & 137.78 & 139.72 \\
        \bottomrule
    \end{tabular}
    \label{tab:kms_table}
\end{table}

The dataset is formed by 12 samples of infrared sky images and their respective labels. The evolution of clouds in the atmosphere is a continuous process. Therefore, the images in the dataset are ordered chronologically to avoid including future information about the state of the system (which will be not available during the implementation of the algorithm). In addition, the samples belong to different days in each of the four seasons to develop a global segmentation method valid for all the seasons in a year. The pixels were manually labelled as either clear-sky $y_{i,j} = -1$ or cloudy $y_{i,j} = 1$. 7 of the images (earlier dates) are used for training and the remaining 5 (later dates) are used for testing. The training dataset (33,600 pixels) have 5 images featuring: contrail, cumulus, altocumulus, clear-sky and altostratus. The testing dataset images (24,000 pixels) show: stratocumulus and cumulus, stratocumulus, cirrocumulus and stratocumulus, altocumulus, cumulus, nimbus and clear-sky

\begin{figure}[!htbp]
    \centering
    \begin{subfigure}{0.325\linewidth}
        \includegraphics[scale = .35, trim = 4 4 4 4, clip]{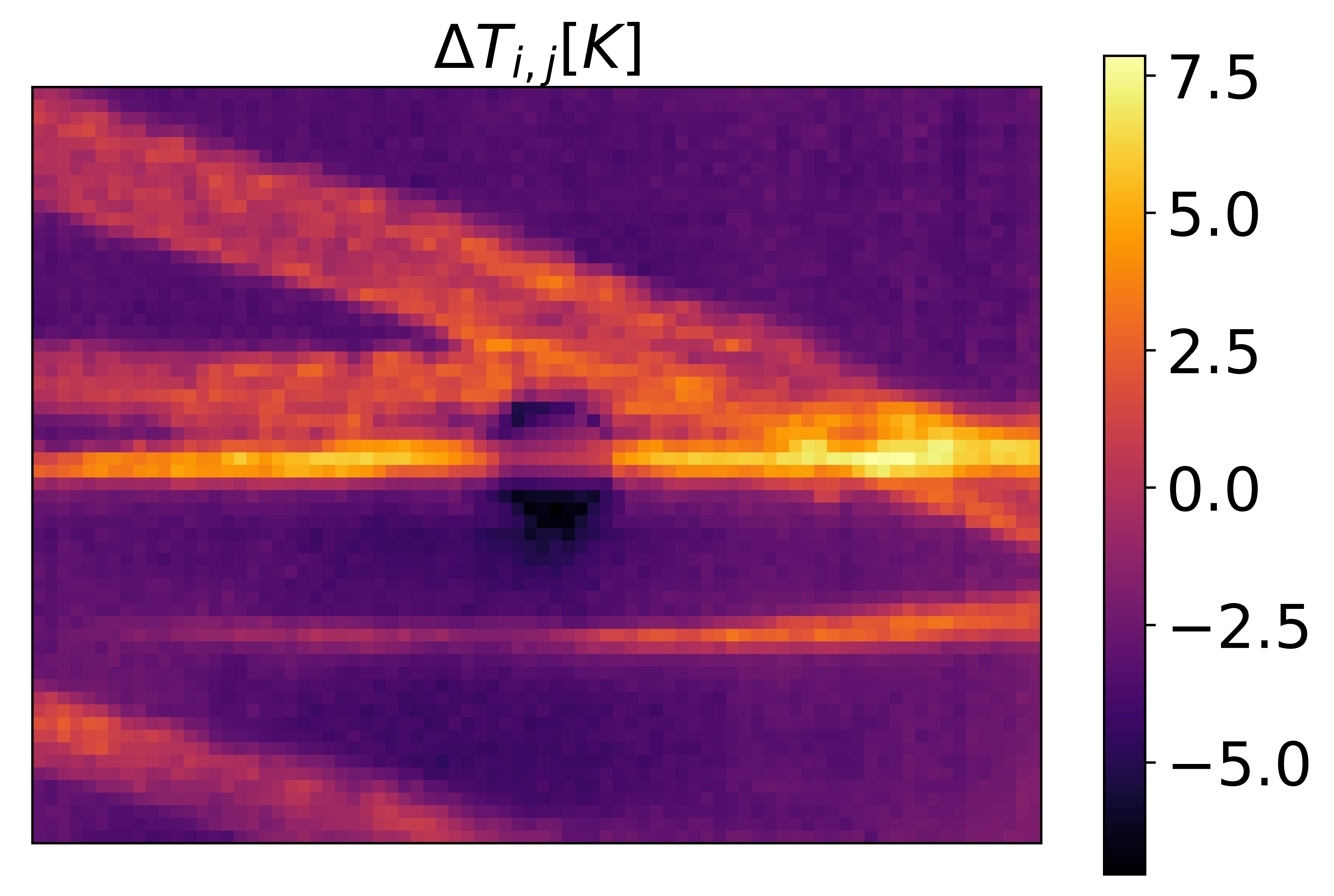}
    \end{subfigure}
    \begin{subfigure}{0.325\linewidth}
        \includegraphics[scale = .35, trim = 4 4 4 4, clip]{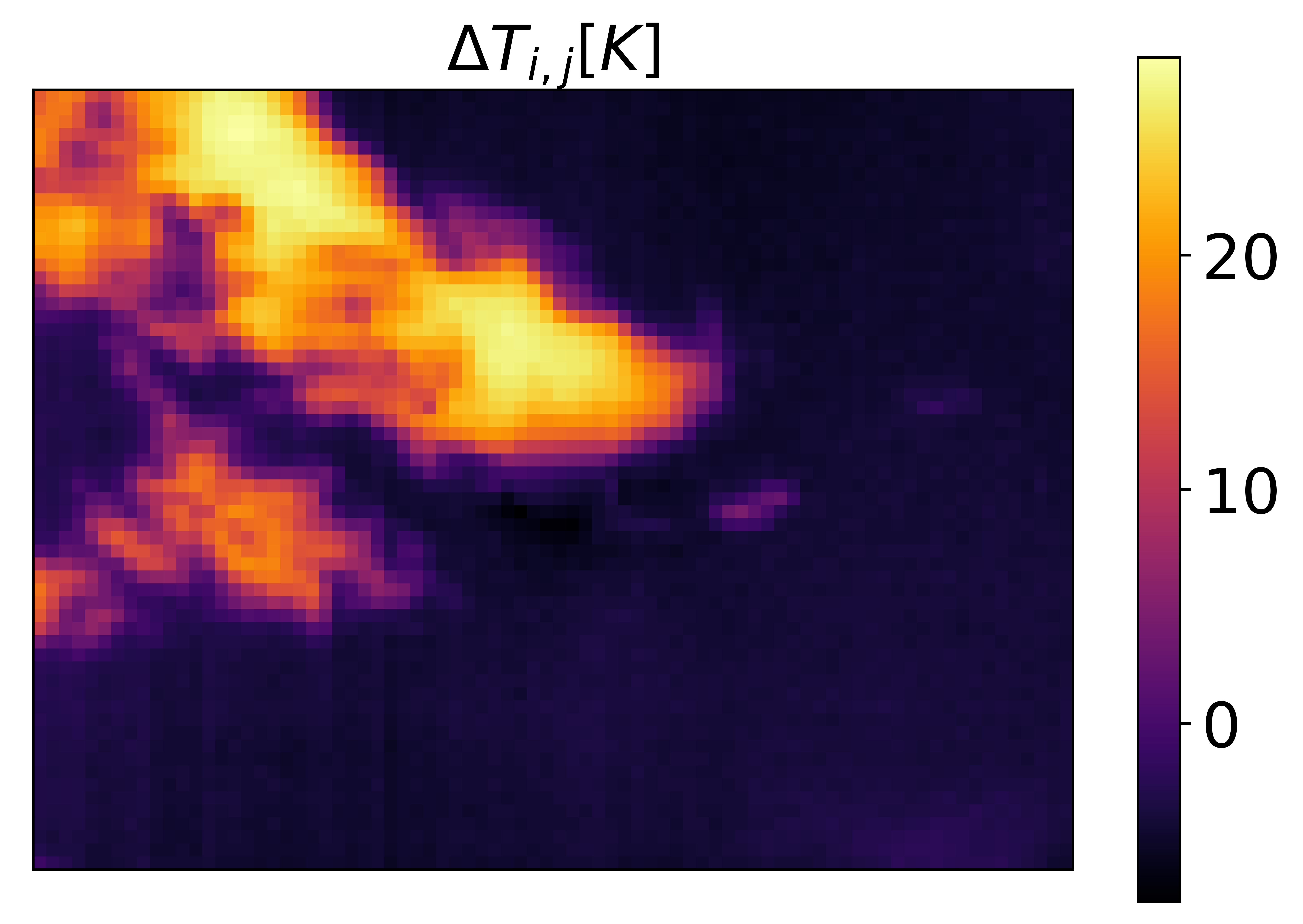}
    \end{subfigure}
    \begin{subfigure}{0.325\linewidth}
        \includegraphics[scale = .35, trim = 4 4 4 4, clip]{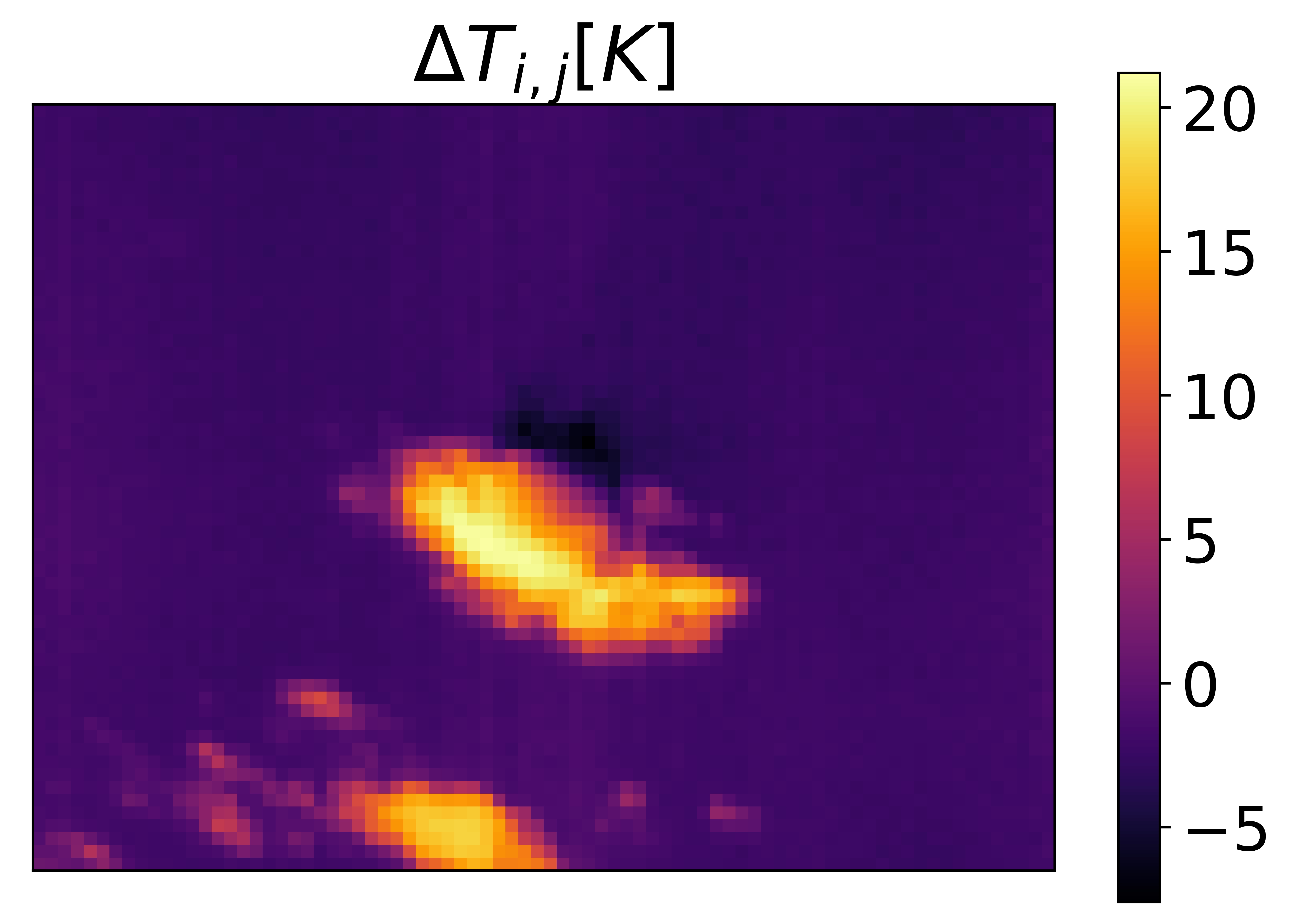}
    \end{subfigure}
    \begin{subfigure}{0.325\linewidth}
        \includegraphics[scale = .35, trim = 4 4 4 4, clip]{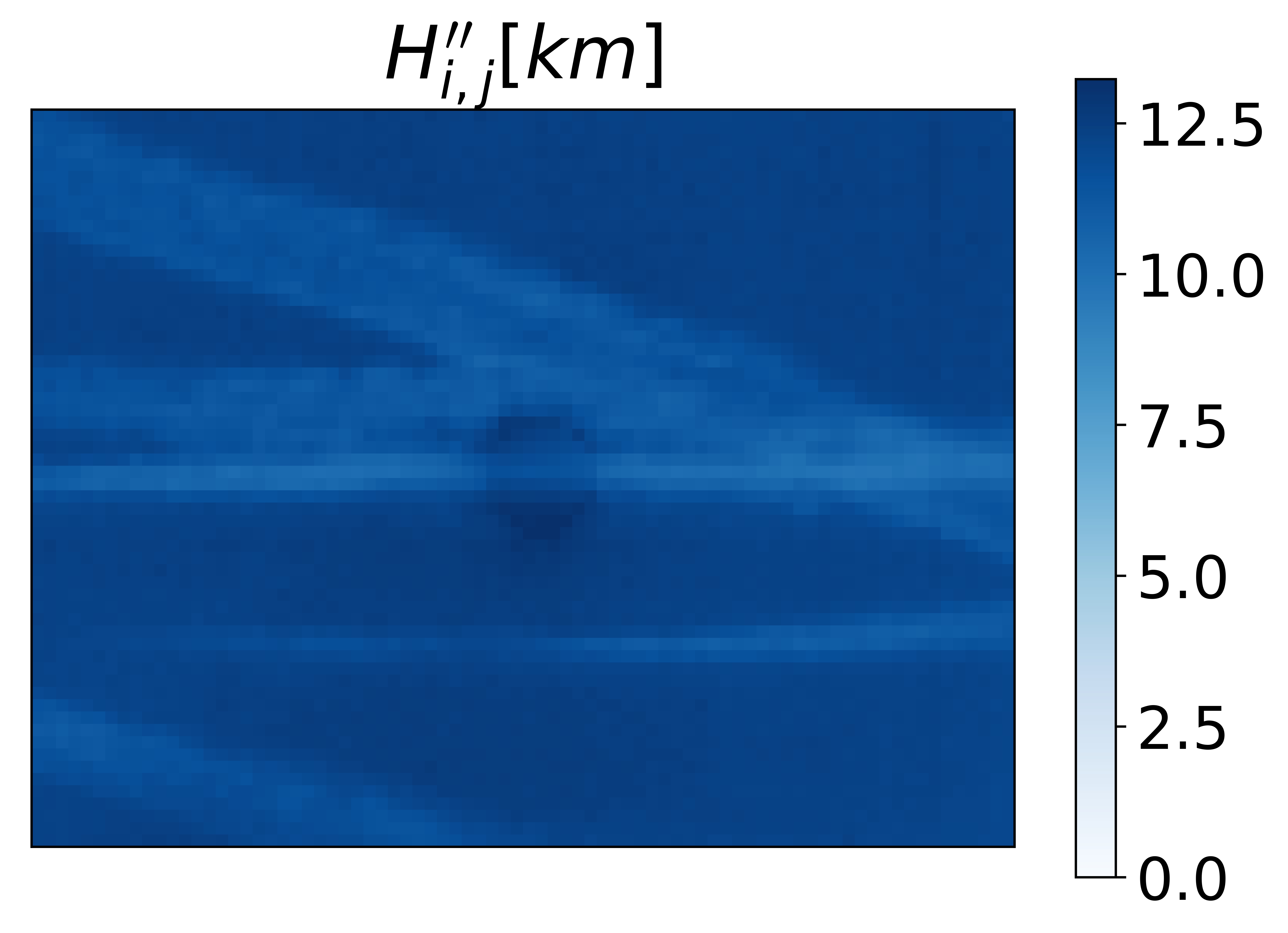}
    \end{subfigure}
    \begin{subfigure}{0.325\linewidth}
        \includegraphics[scale = .35, trim = 4 4 4 4, clip]{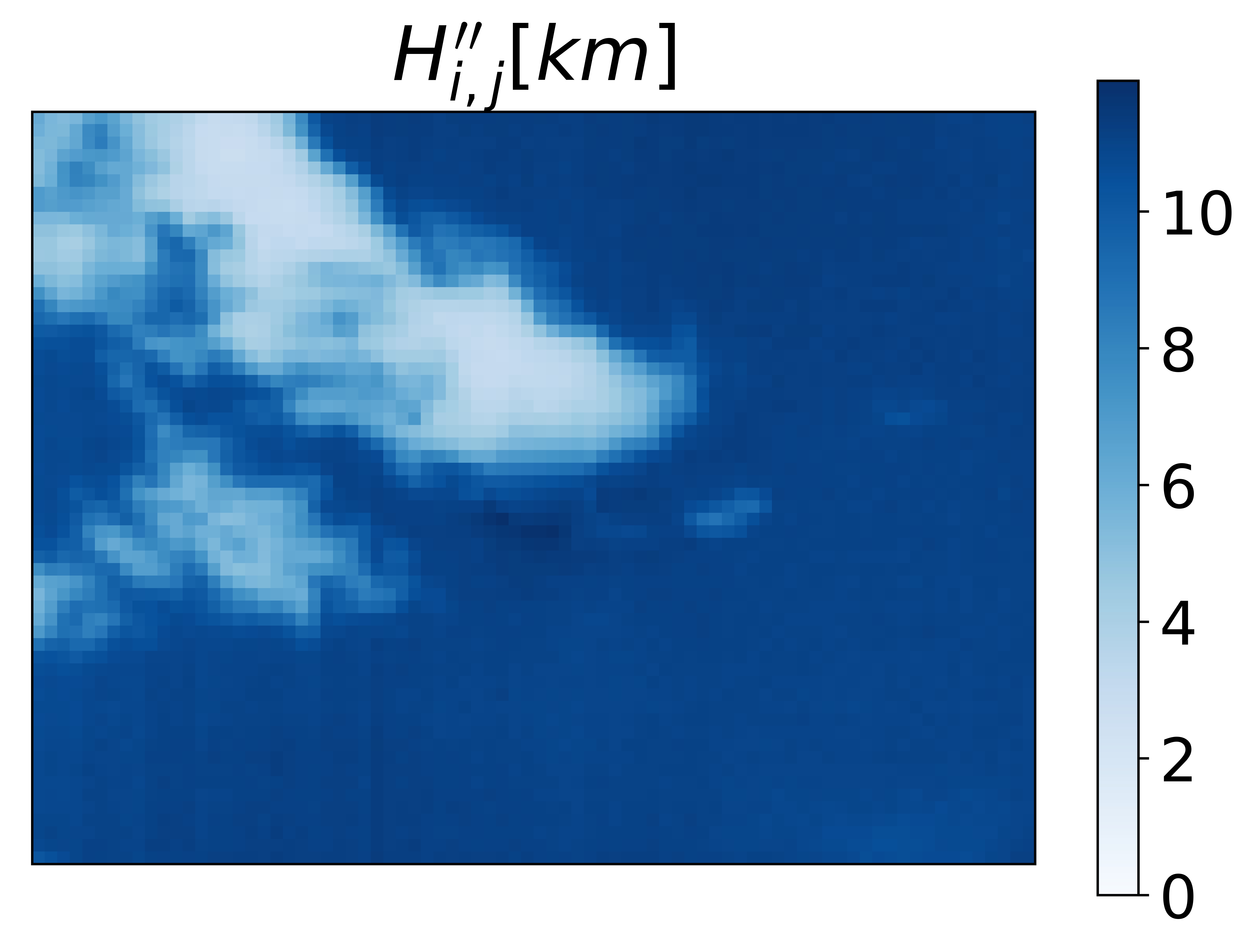}
    \end{subfigure}
    \begin{subfigure}{0.325\linewidth}
        \includegraphics[scale = .35, trim = 4 4 4 4, clip]{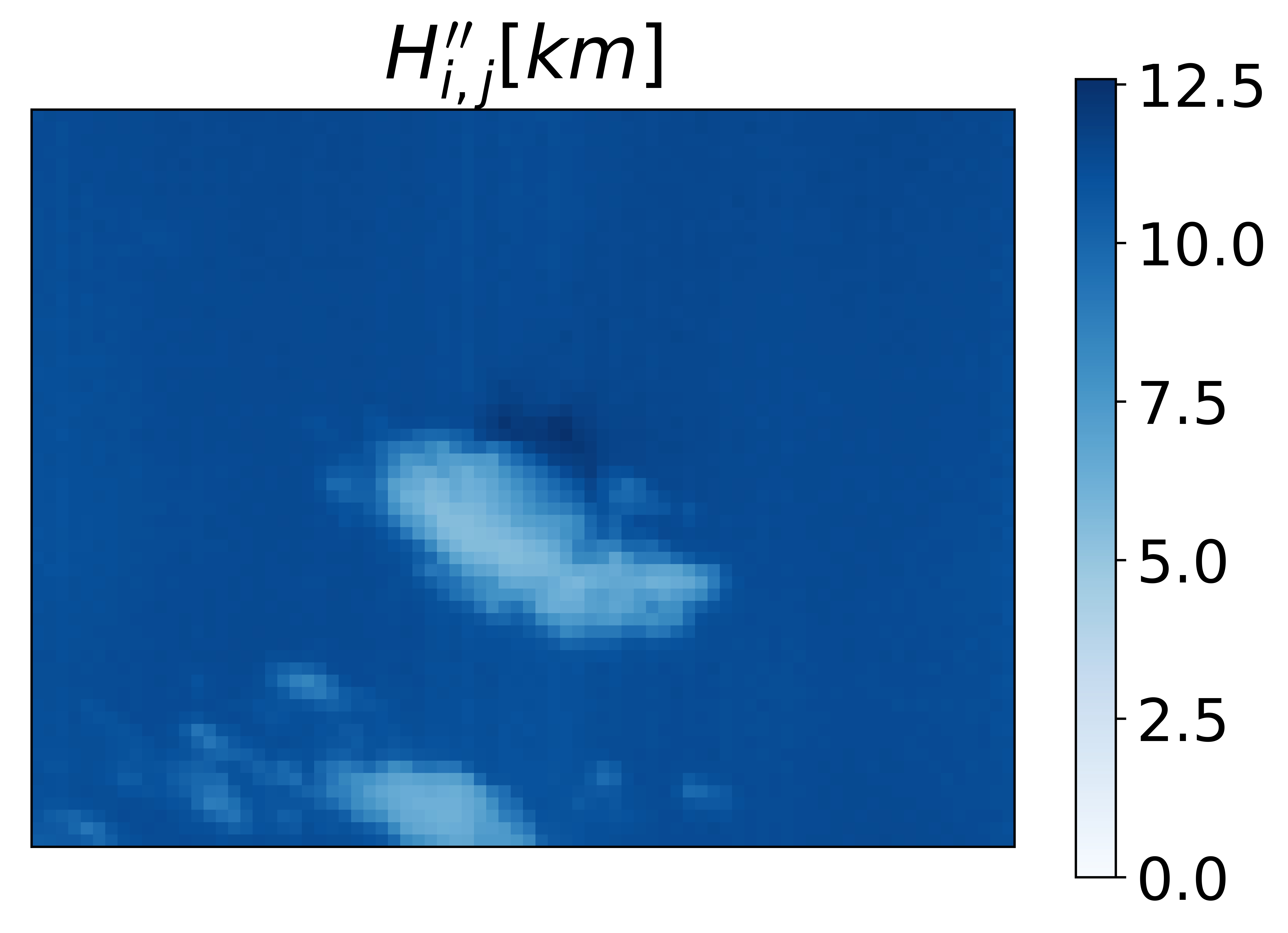}
    \end{subfigure}
    \begin{subfigure}{0.325\linewidth}
        \includegraphics[scale = .35, trim = 4 4 4 4, clip]{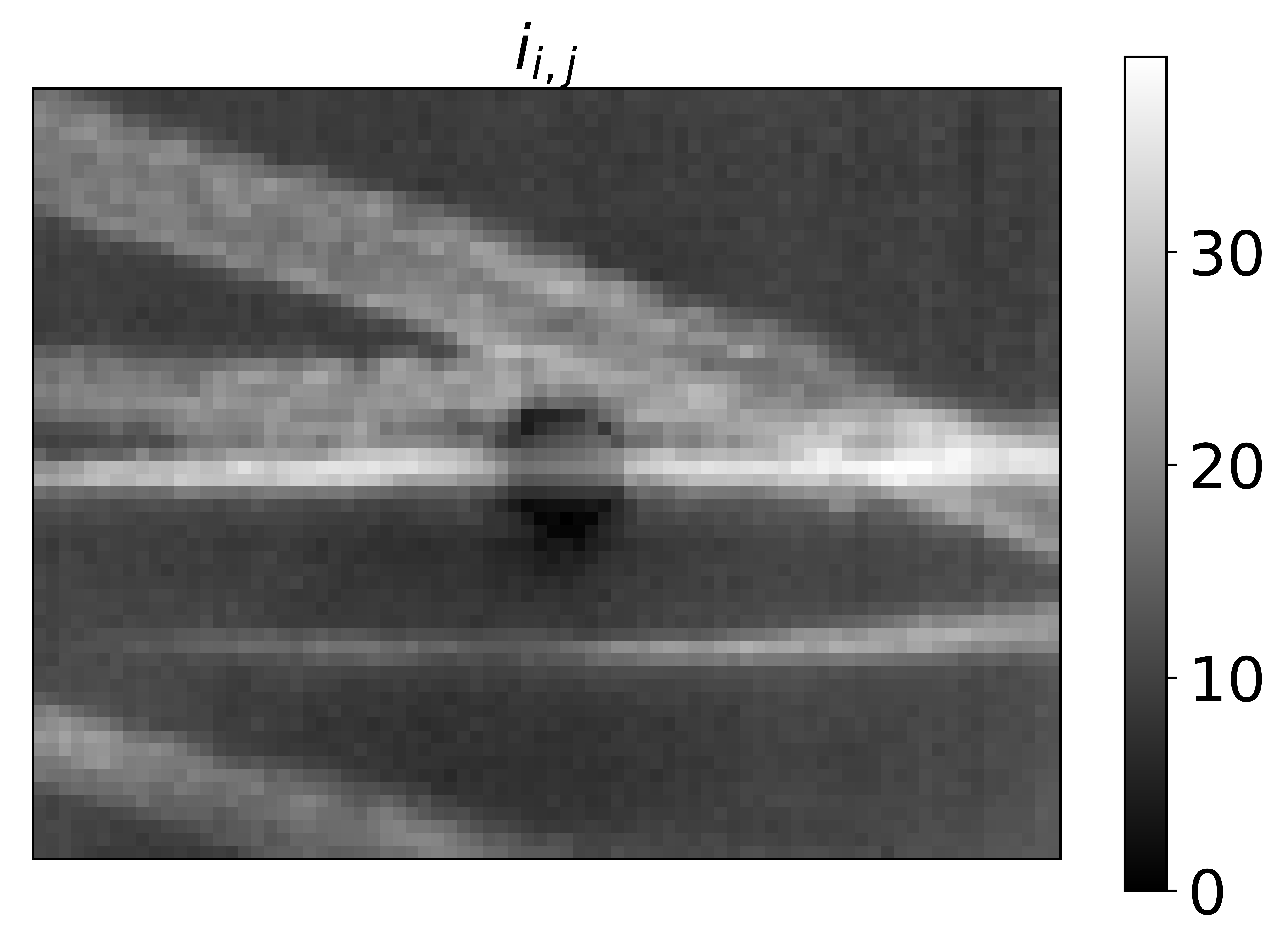}
    \end{subfigure}
    \begin{subfigure}{0.325\linewidth}
        \includegraphics[scale = .35, trim = 4 4 4 4, clip]{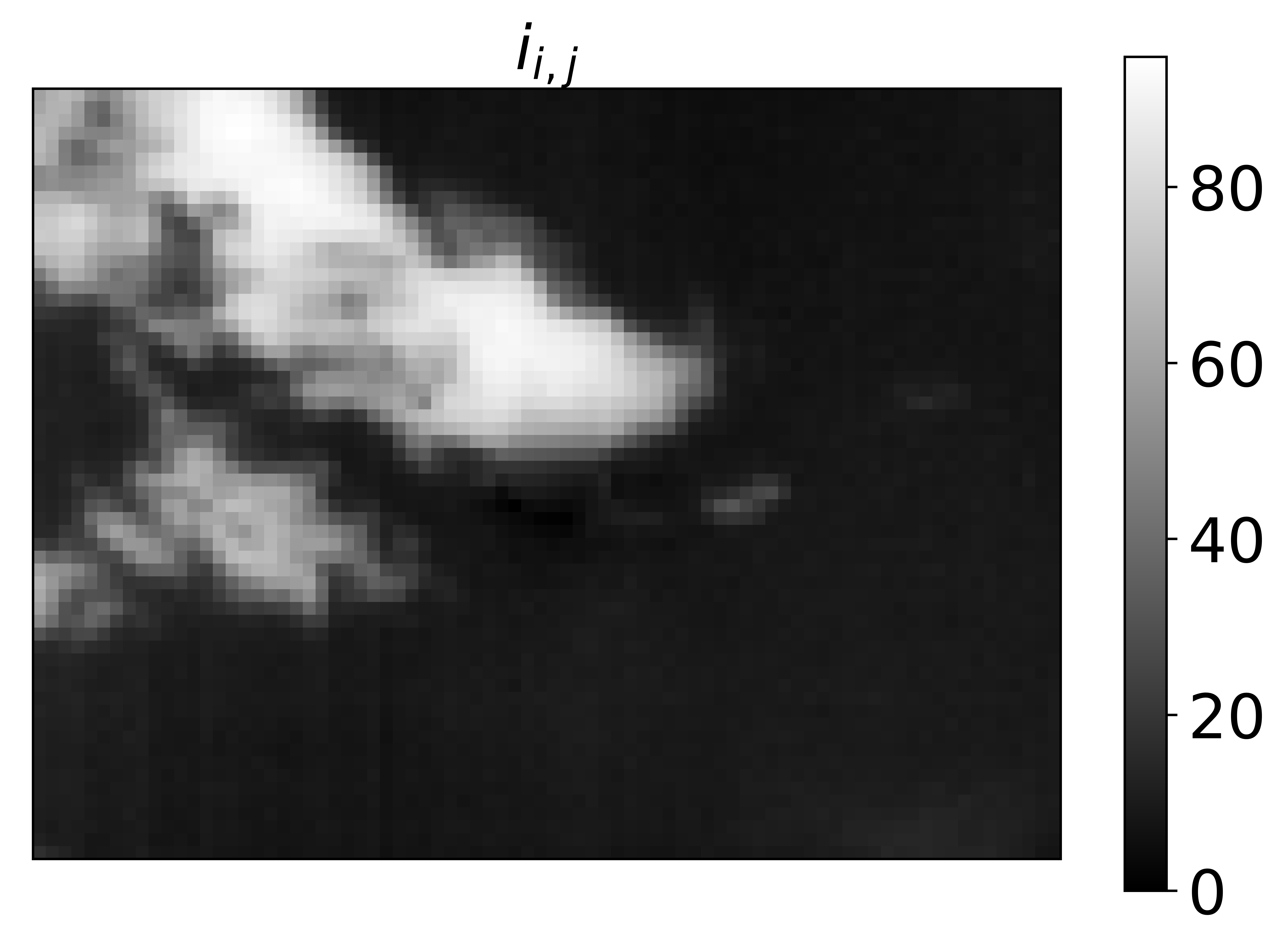}
    \end{subfigure}
    \begin{subfigure}{0.325\linewidth}
        \includegraphics[scale = .35, trim = 4 4 4 4, clip]{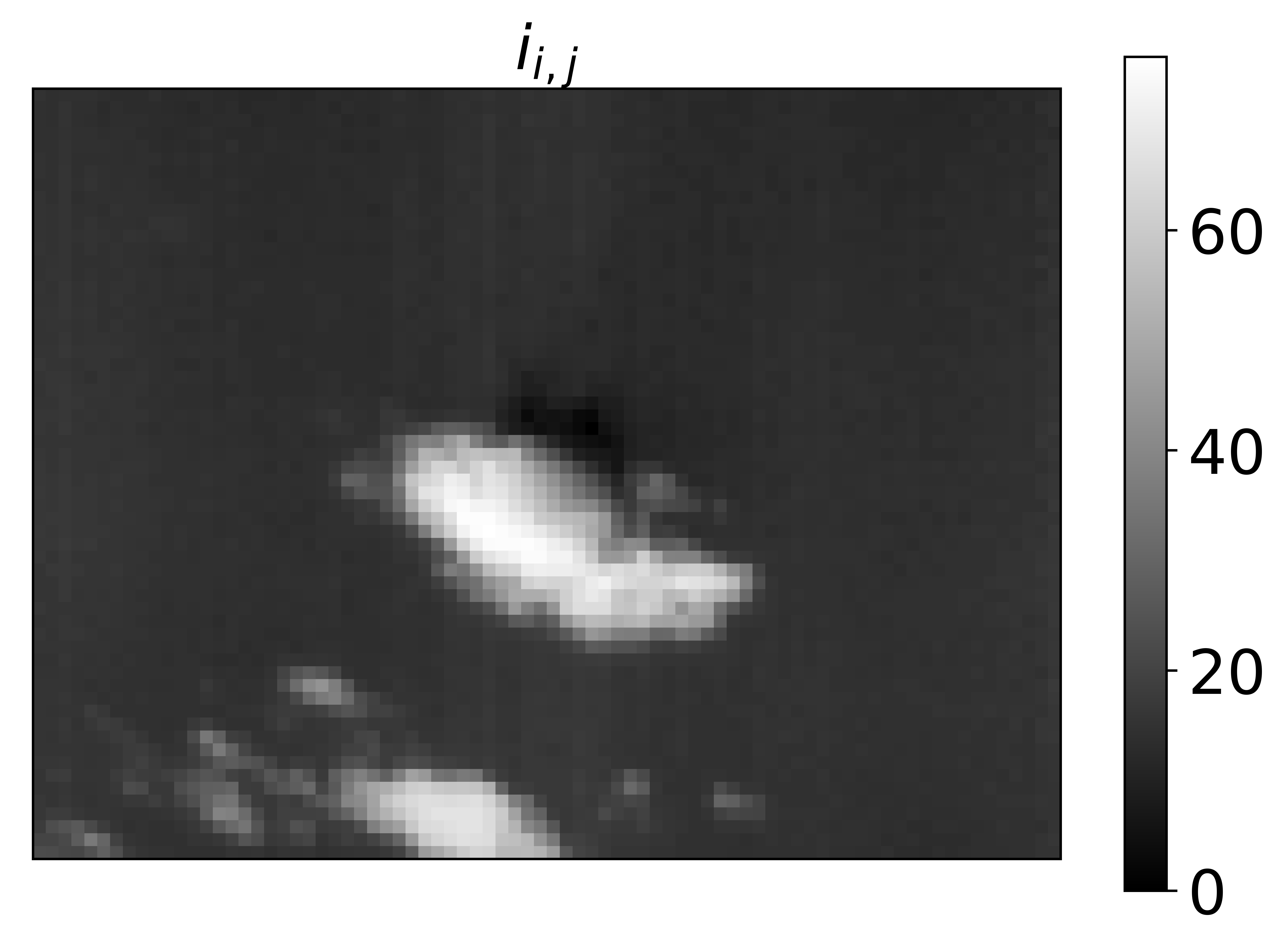}
    \end{subfigure}
    \begin{subfigure}{0.325\linewidth}
        \includegraphics[scale = .35, trim = 4 4 4 4, clip]{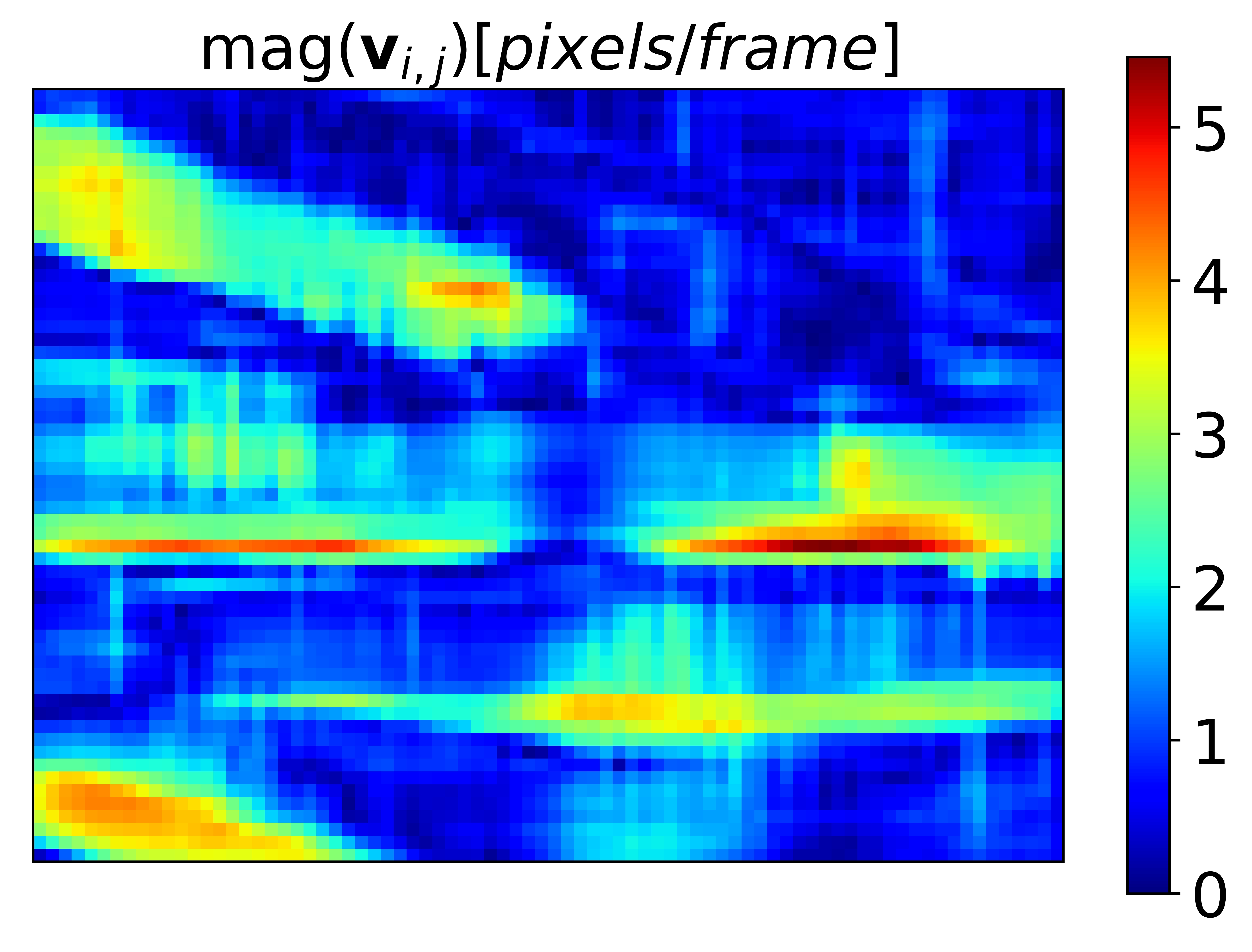}
    \end{subfigure}
    \begin{subfigure}{0.325\linewidth}
        \includegraphics[scale = .35, trim = 4 4 4 4, clip]{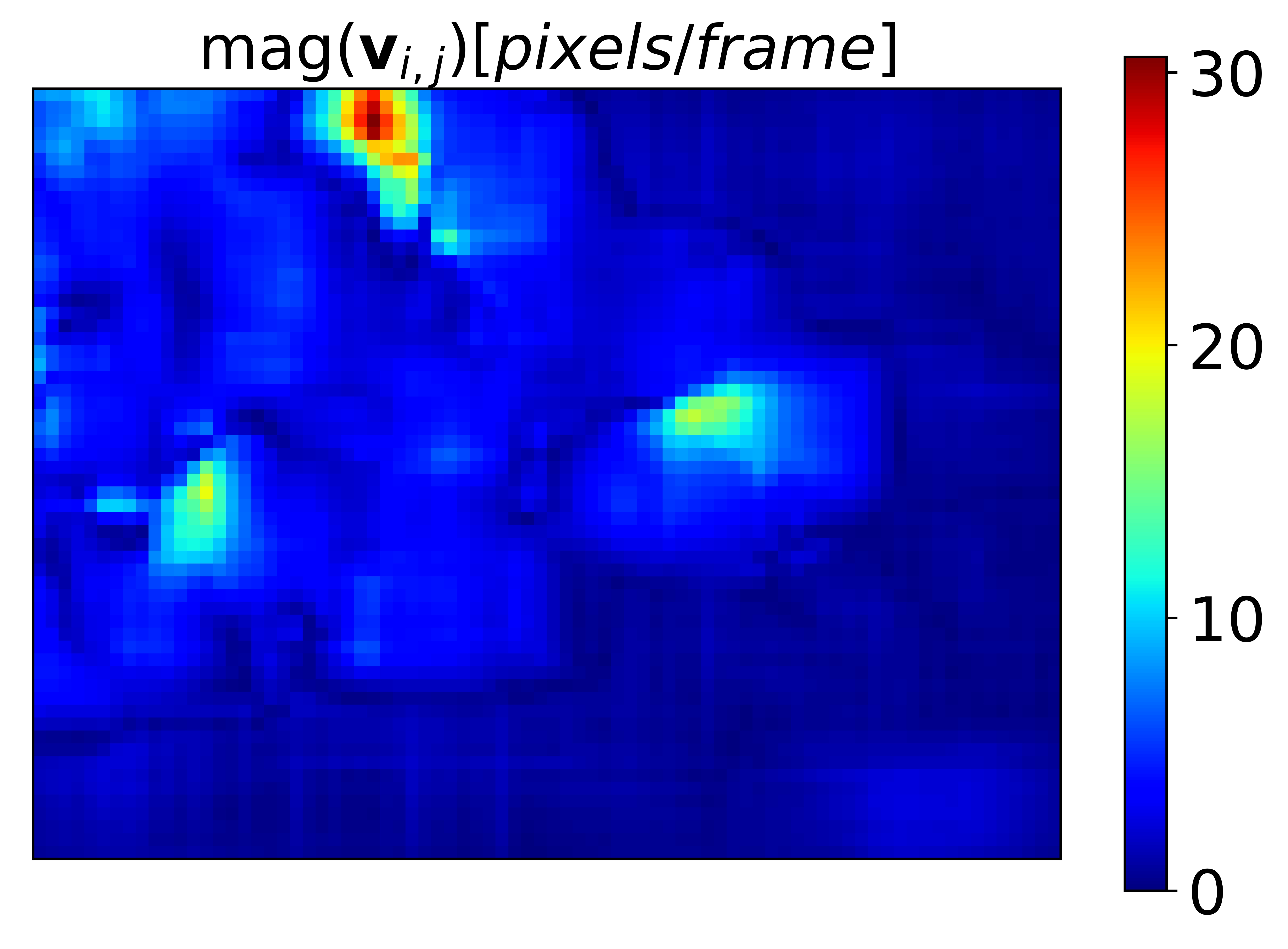}
    \end{subfigure}
    \begin{subfigure}{0.325\linewidth}
        \includegraphics[scale = .35, trim = 4 4 4 4, clip]{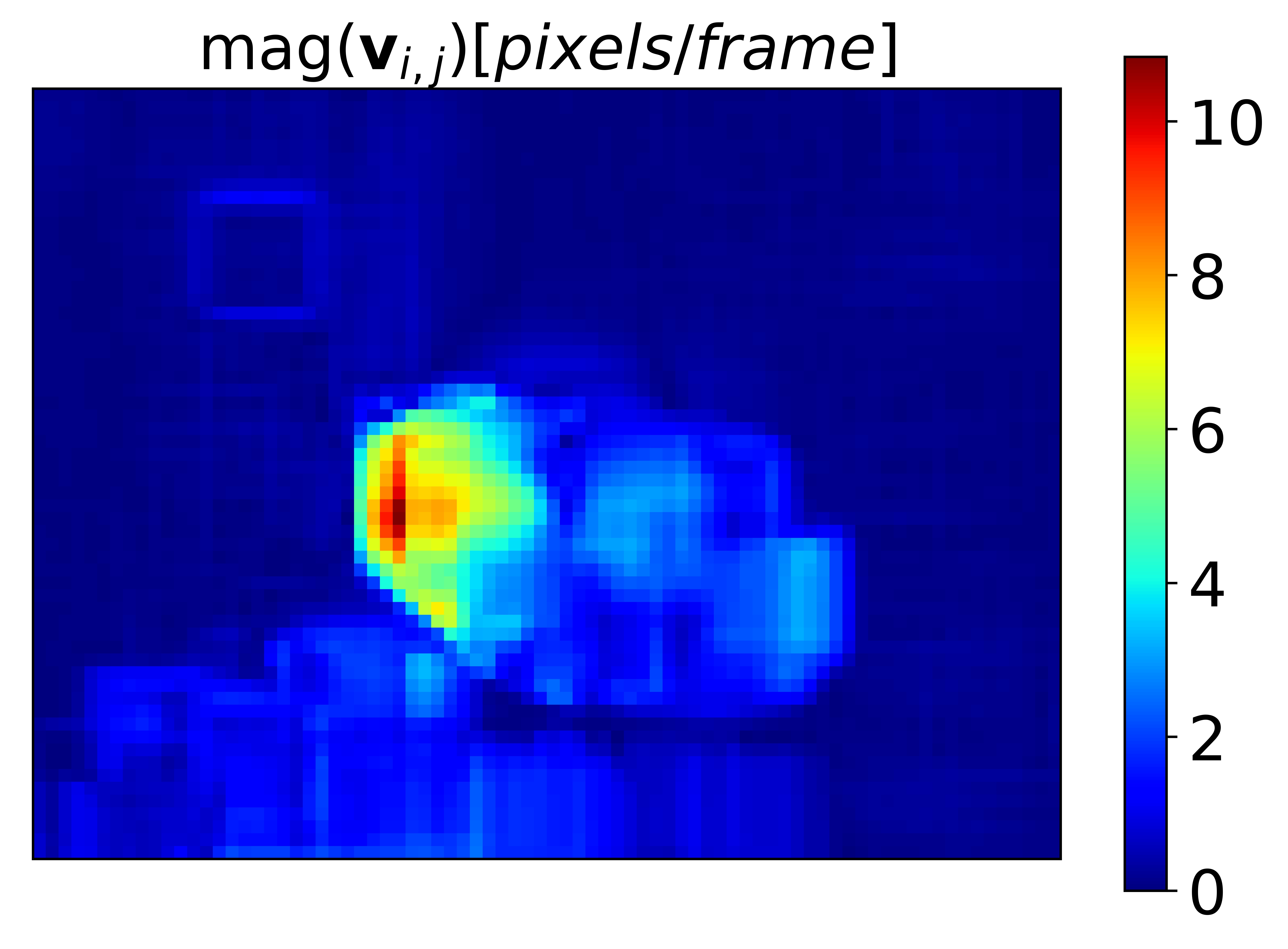}
    \end{subfigure}
    \caption{This figure shows the features extracted from three testing images. The image in the first row show the increments of temperature with respect to the height of the Tropopause. The features in the second row show the heights of the clouds. The images in the third row show the normalized intensity of the pixels. The images in the fourth row show the magnitude of the velocity vectors. }
    \label{fig:features}
\end{figure}

Leave-One-Out Cross-Validation (LOO-CV) is applied to validate the hyperparameter of the models and the prior $\lambda$ in Eq. \eqref{eq:virtual_prior}. In each iteration of the LOO-CV routine an image is left out for validation and the remaining 6 images are used for training the model. A j-statistic is computed for each of the 7 LOO-CV iterations. The validation j-statistic (used as selection criteria of the hyperparameters) is calculated averaging together the j-statistics obtained in each LOO-CV iteration. The selected set of parameters and $\lambda$ are those which yield the maximum validation j-statistic among all the cross-validated combinations of sets of hyperparameters and $\lambda$.

\begin{figure}[!htbp]
    \centering
    \begin{subfigure}{\linewidth}
        \ \ \includegraphics[scale = .35, trim = 4 4 4 4, clip]{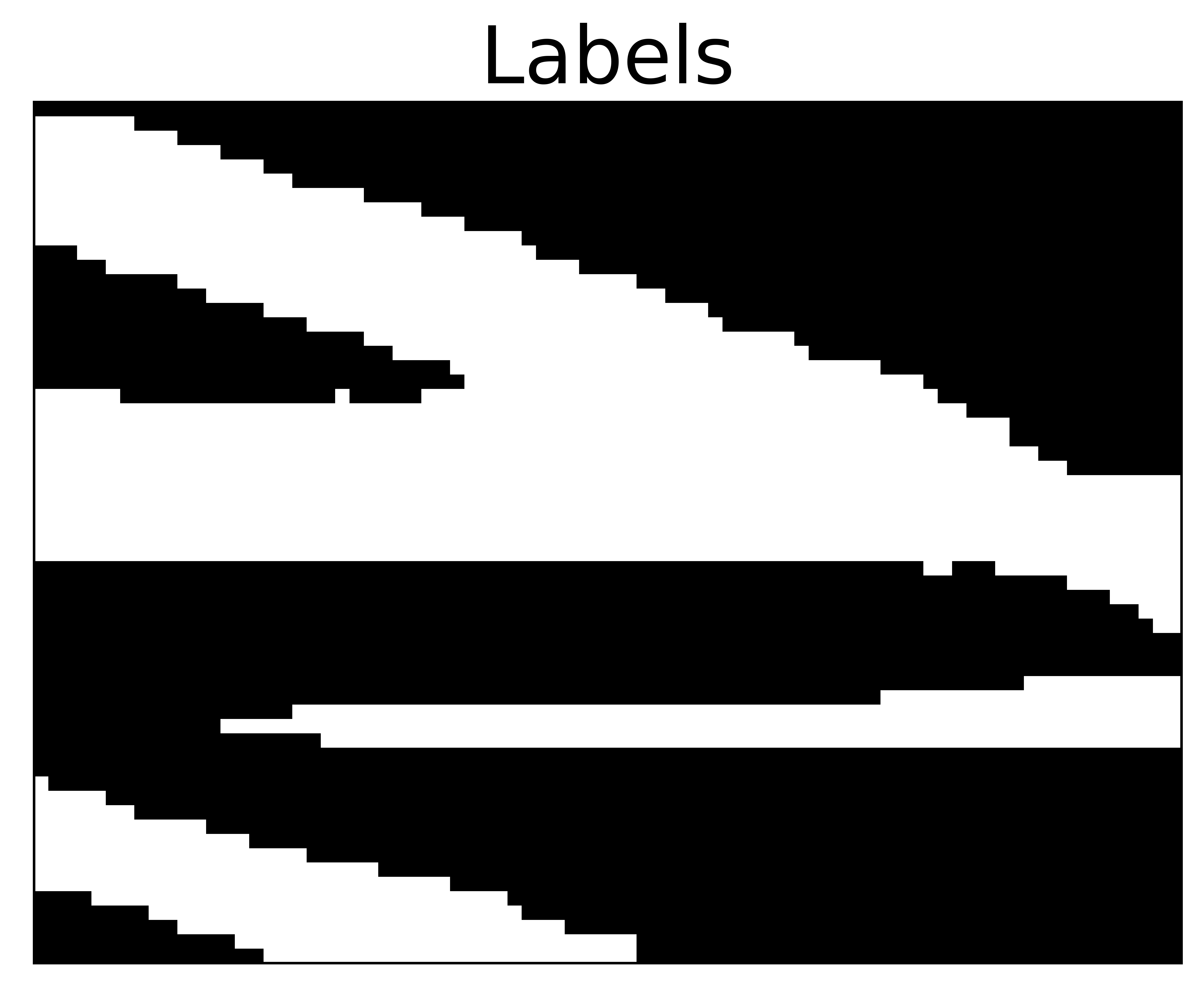}
        \ \ \ \includegraphics[scale = .35, trim = 4 4 4 4, clip]{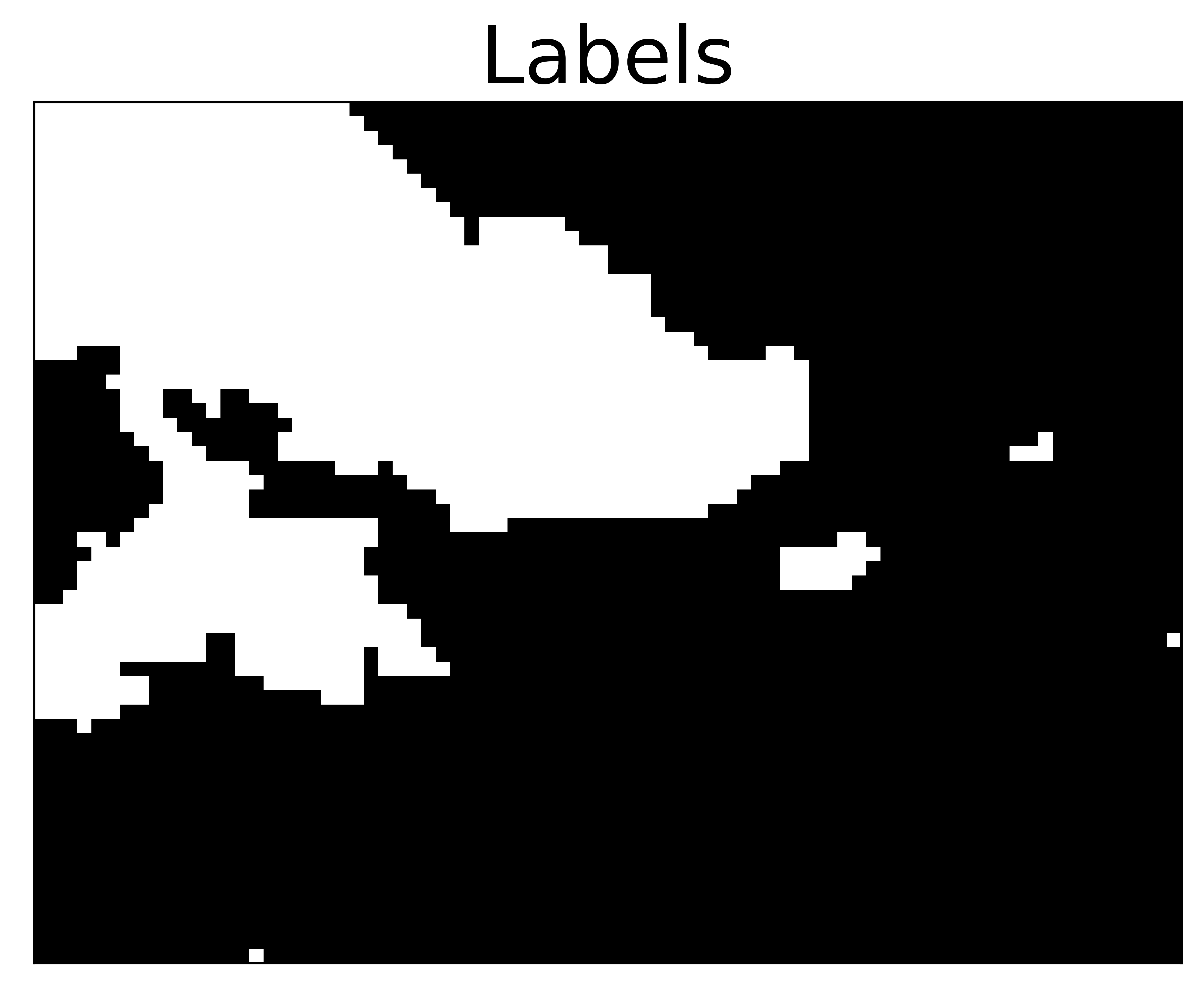}
        \ \ \ \ \ \includegraphics[scale = .35, trim = 4 4 4 4, clip]{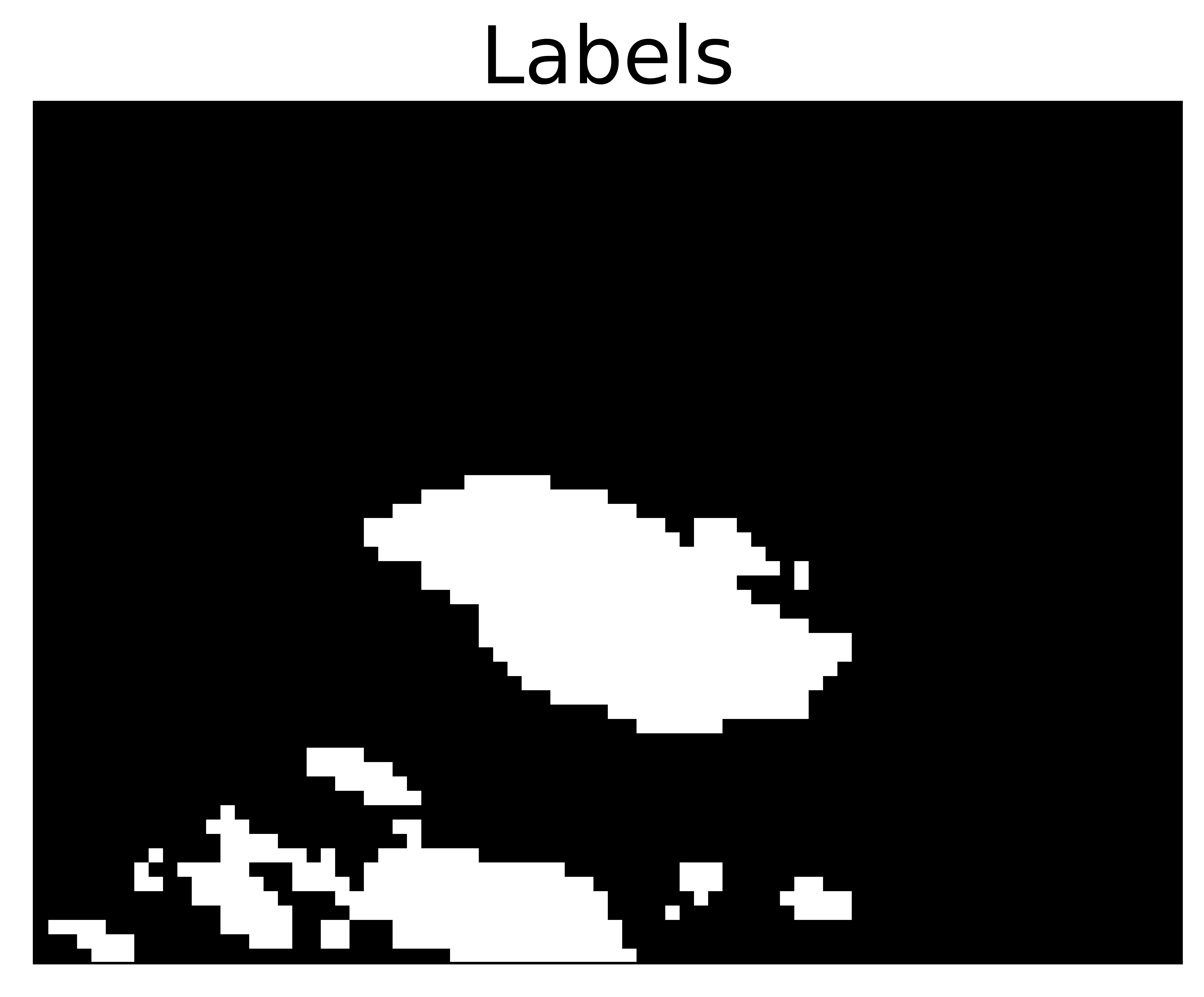}
    \end{subfigure}
    \begin{subfigure}{\linewidth}
        \includegraphics[scale = .35, trim = 4 4 4 4, clip]{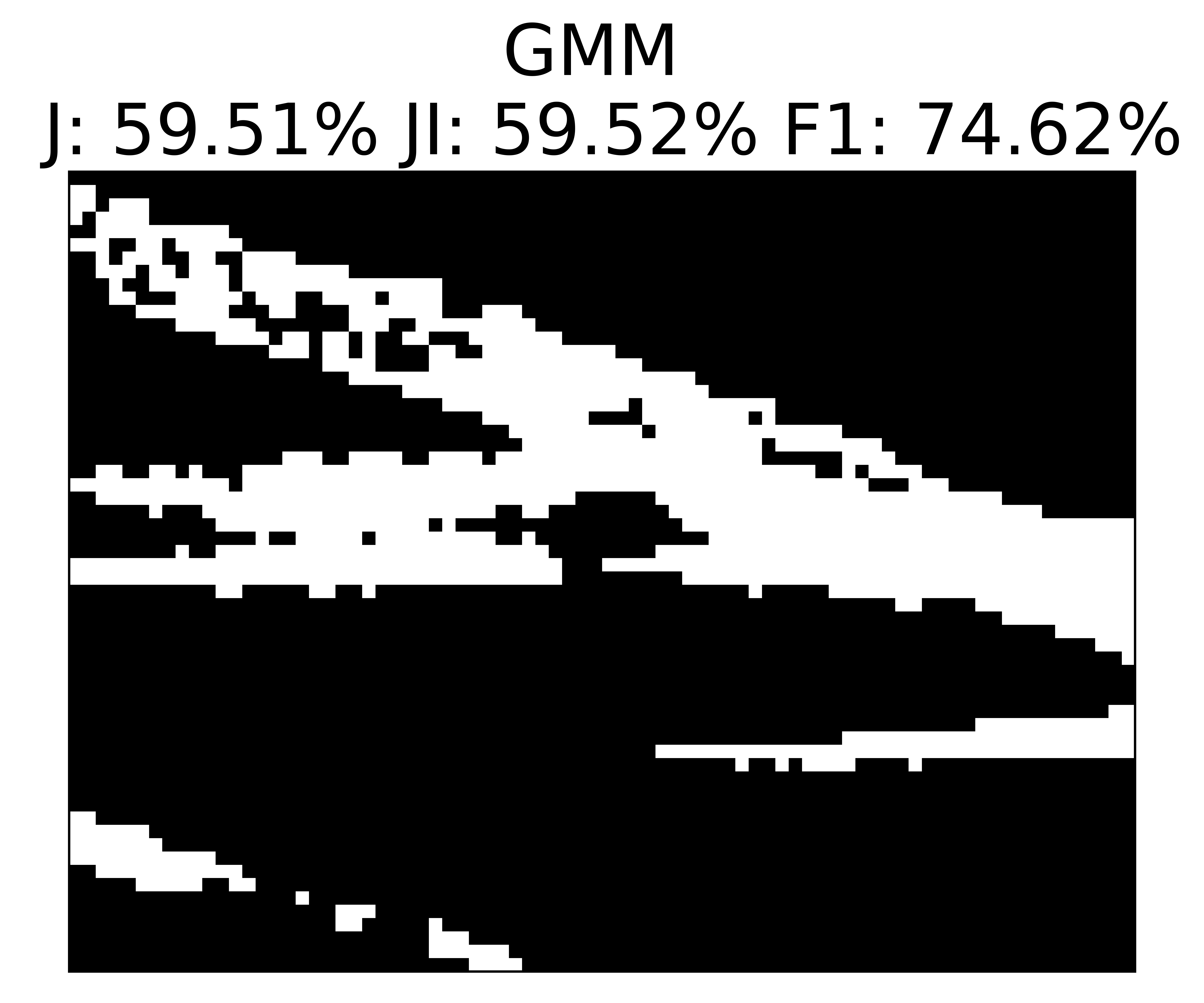}
        \includegraphics[scale = .35, trim = 4 4 4 4, clip]{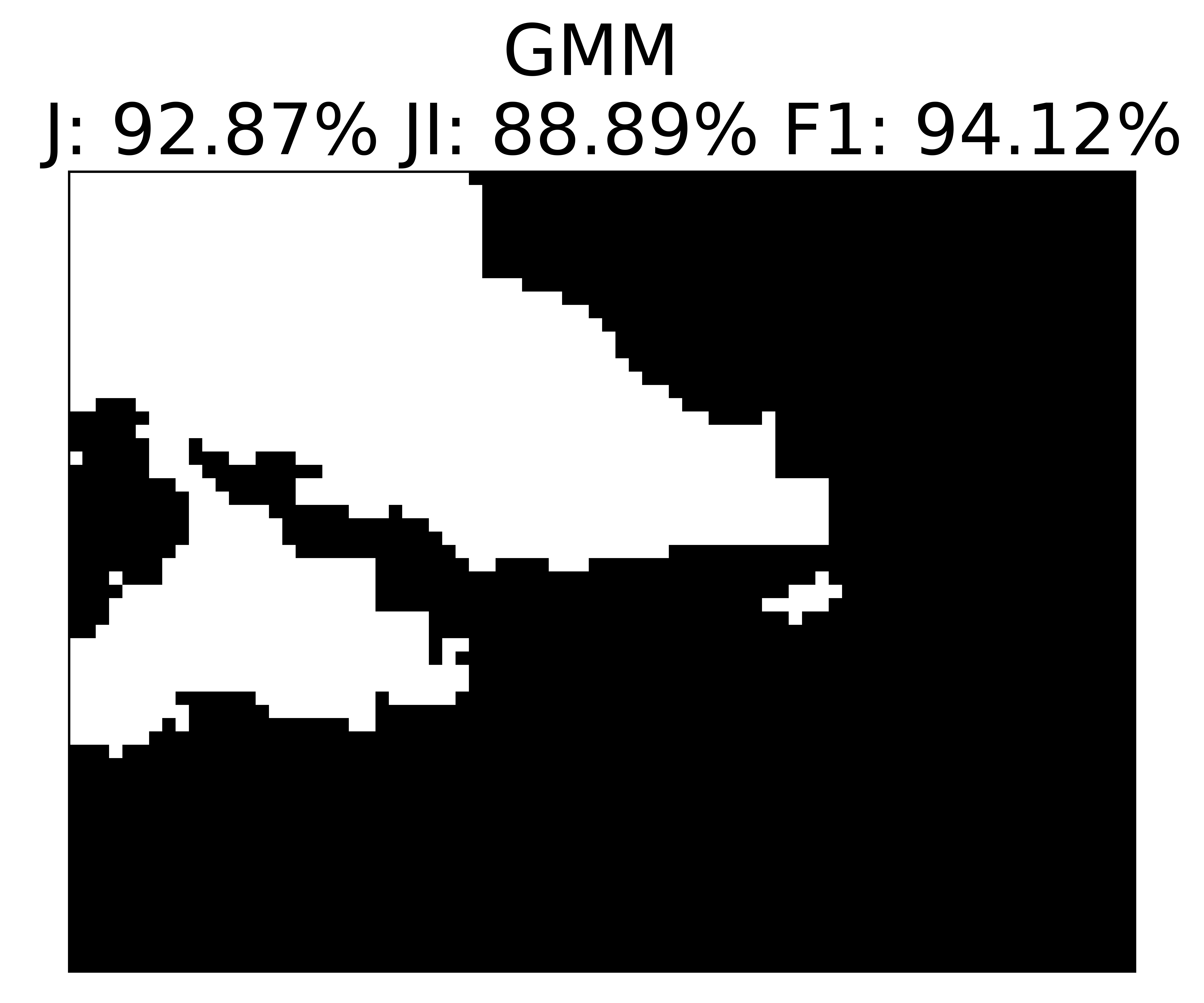}
        \includegraphics[scale = .35, trim = 4 4 4 4, clip]{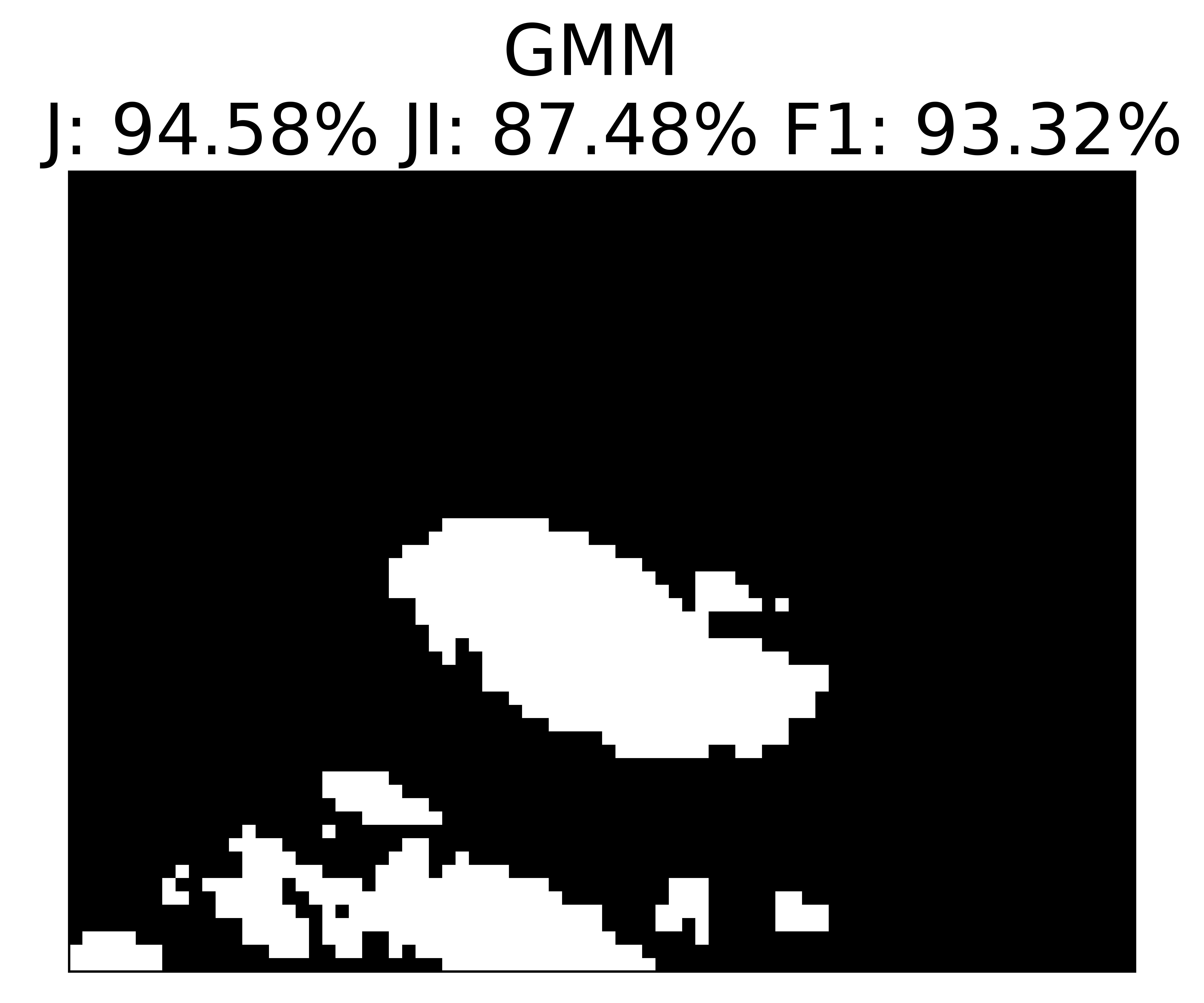}
    \end{subfigure}
    \begin{subfigure}{\linewidth}
        \includegraphics[scale = .35, trim = 4 4 4 4, clip]{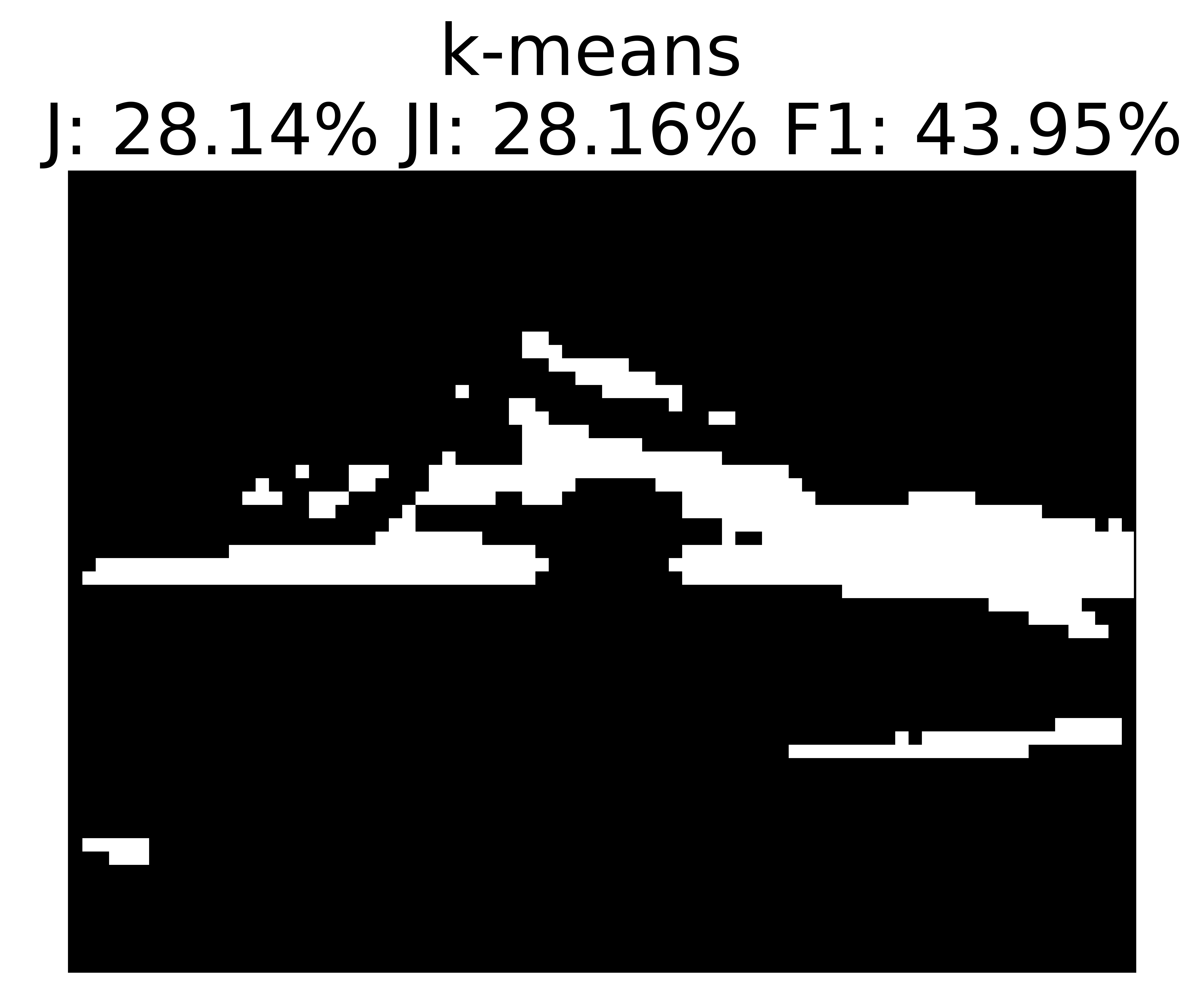}
        \includegraphics[scale = .35, trim = 4 4 4 4, clip]{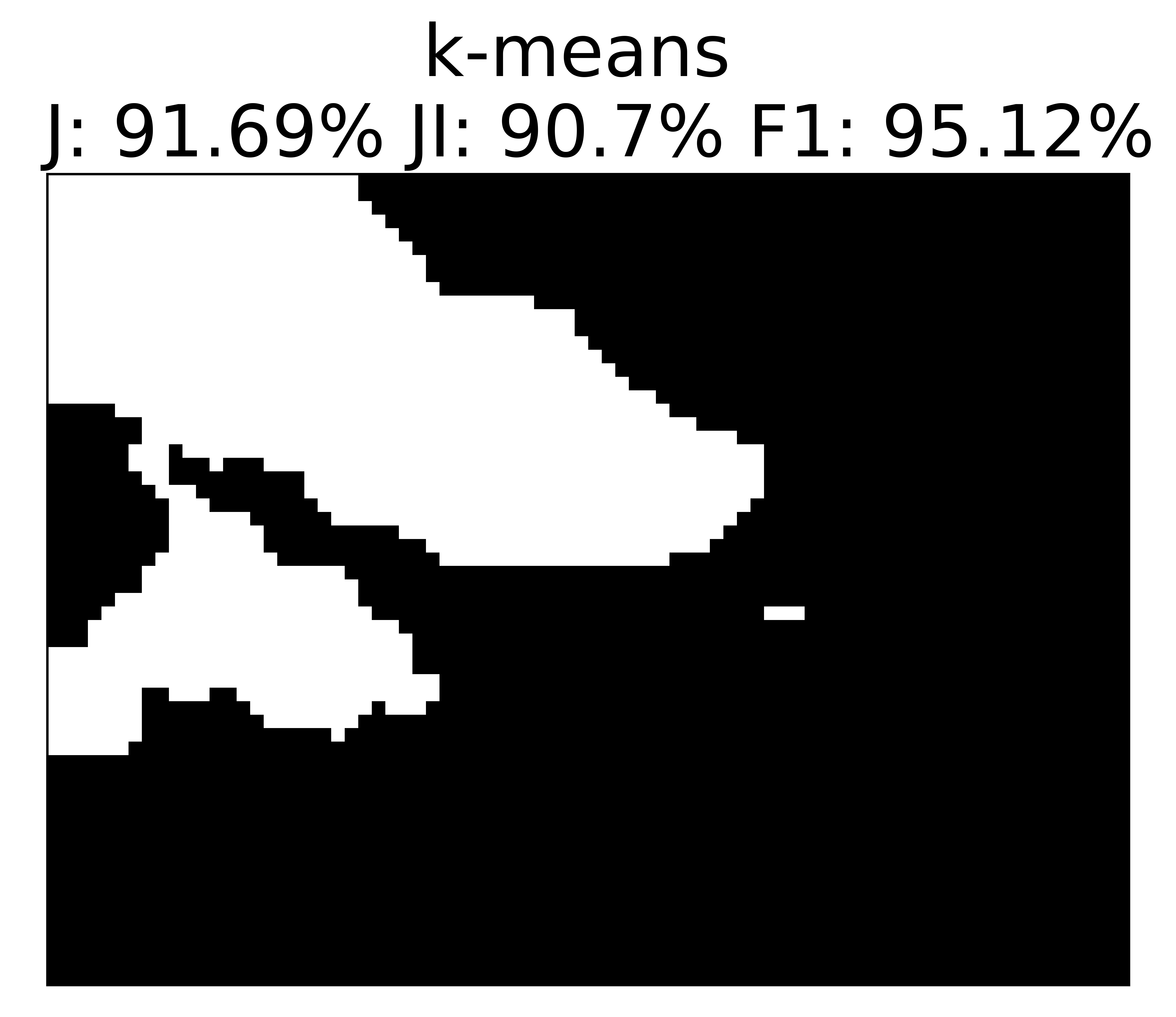}
        \ \includegraphics[scale = .35, trim = 4 4 4 4, clip]{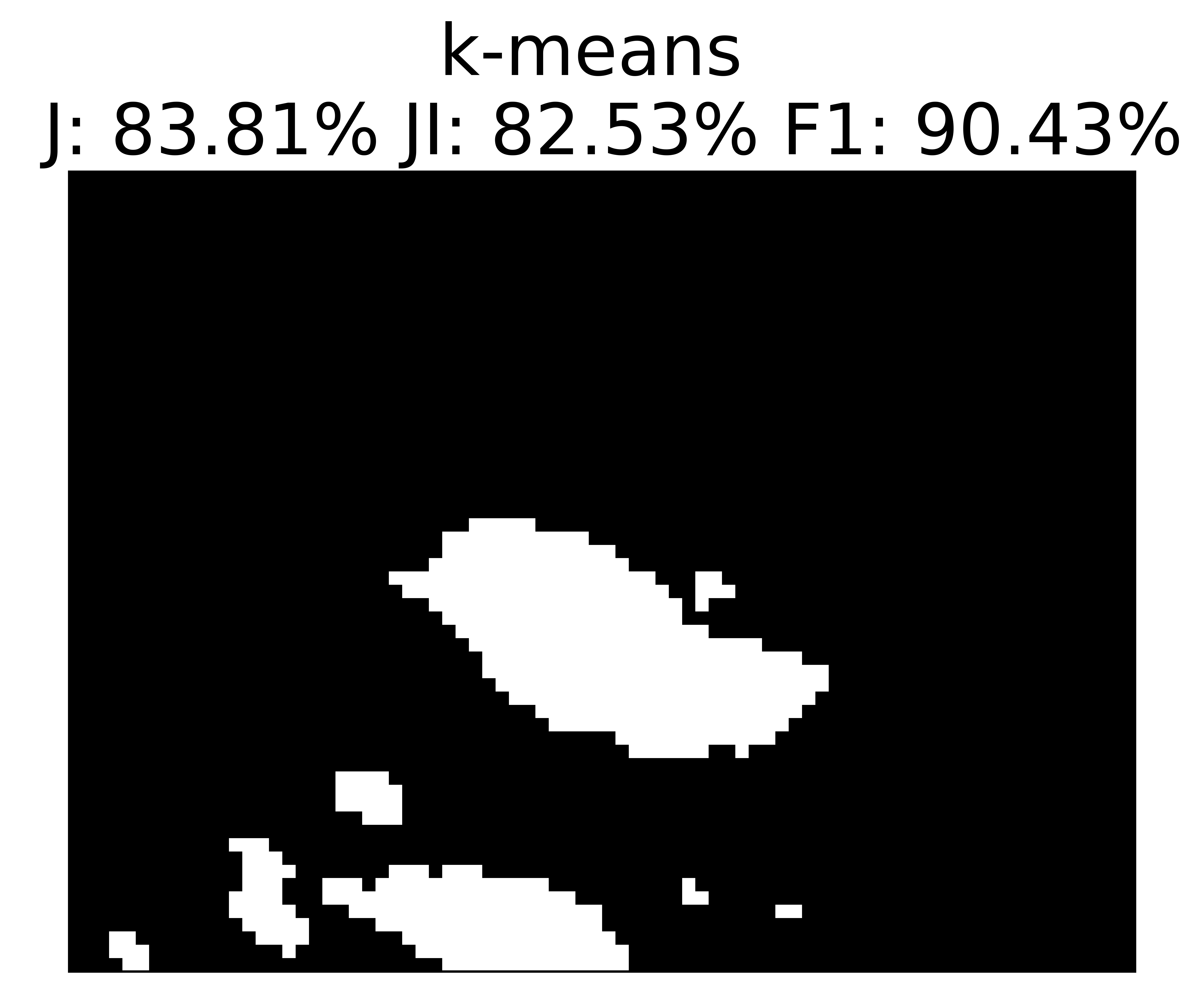}
    \end{subfigure}
    \begin{subfigure}{\linewidth}
        \includegraphics[scale = .35, trim = 4 4 4 4, clip]{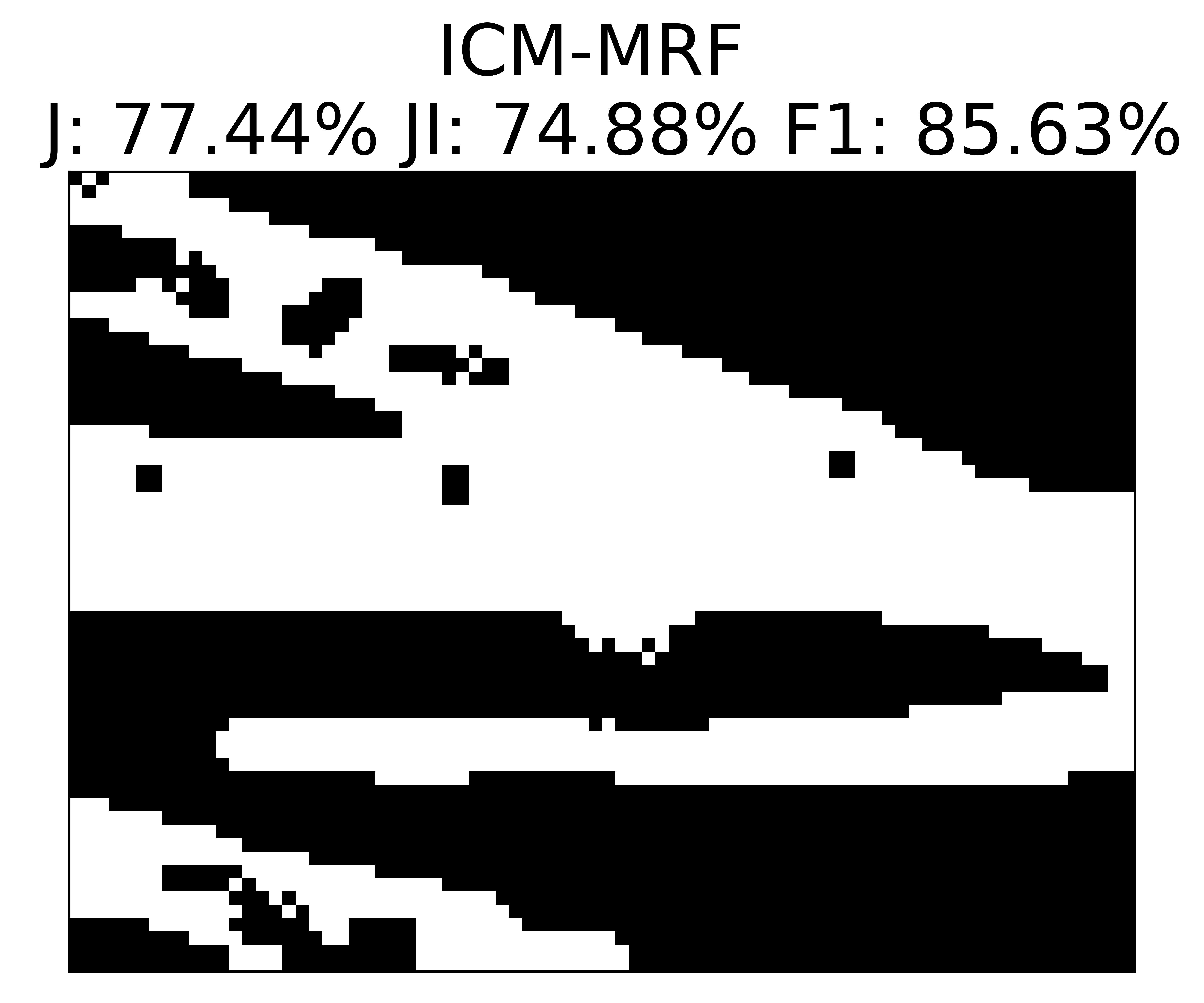}
        \includegraphics[scale = .35, trim = 4 4 4 4, clip]{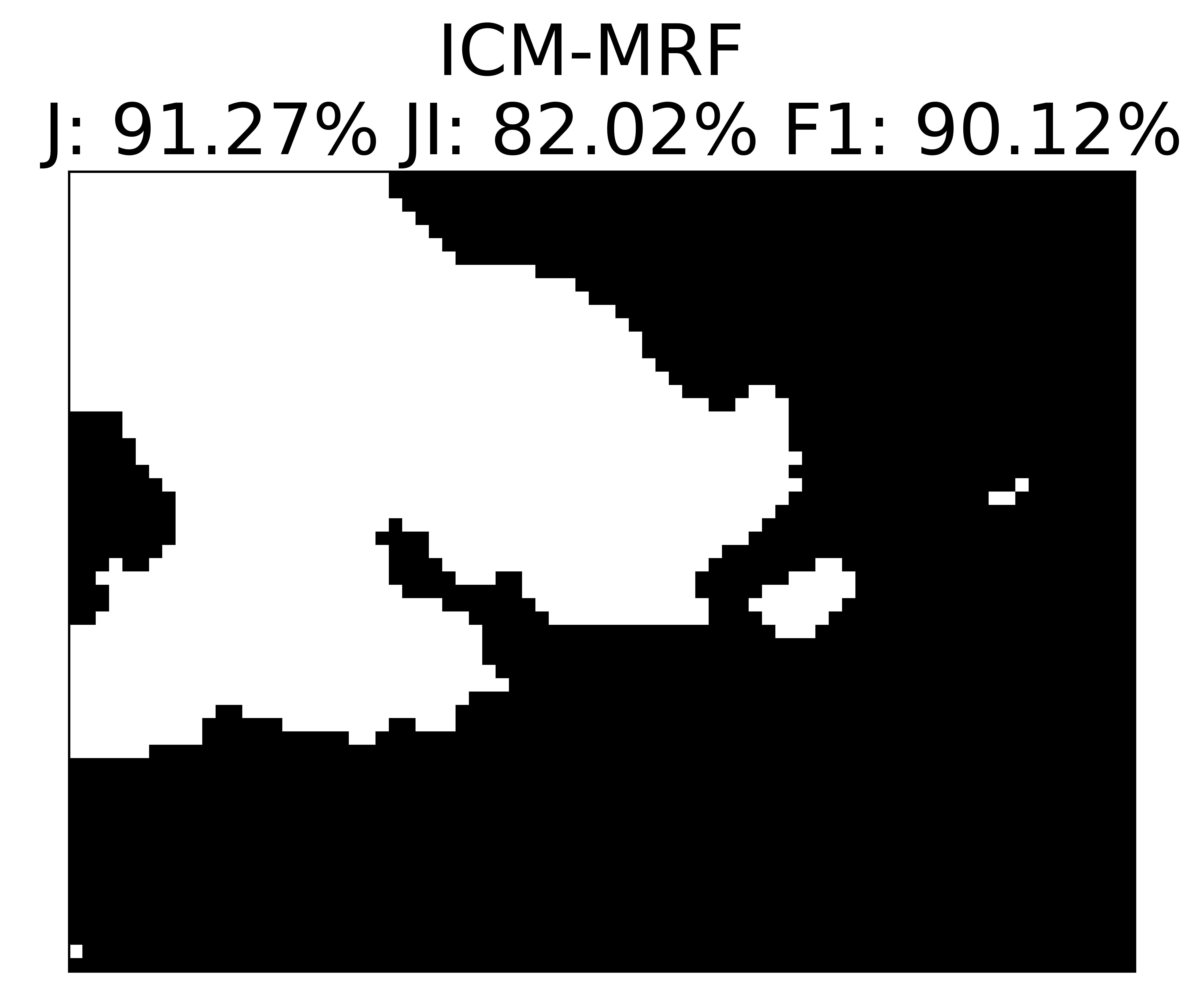}
        \includegraphics[scale = .35, trim = 4 4 4 4, clip]{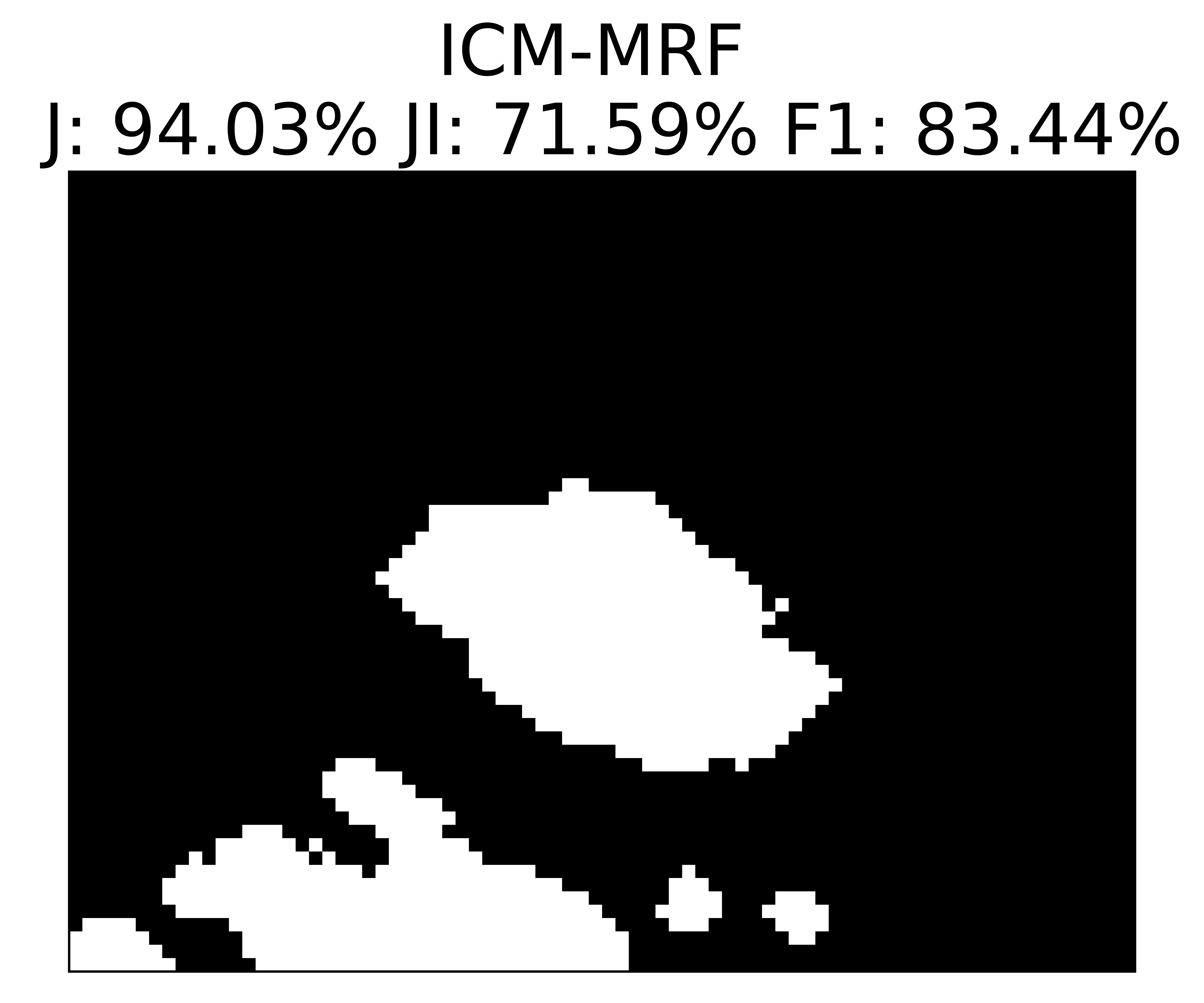}
    \end{subfigure}
    \begin{subfigure}{\linewidth}
        \includegraphics[scale = .35, trim = 4 4 4 4, clip]{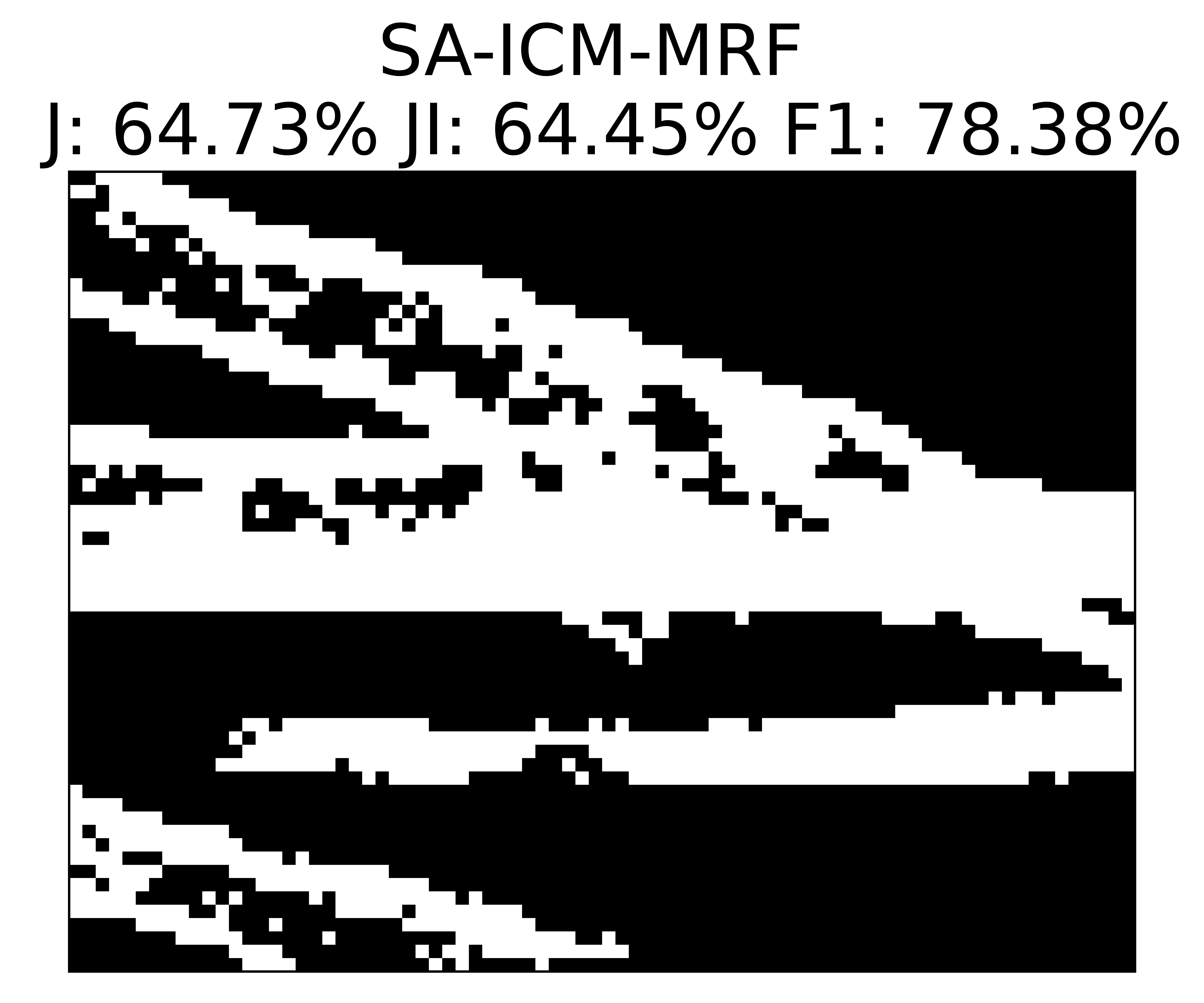}
        \includegraphics[scale = .35, trim = 4 4 4 4, clip]{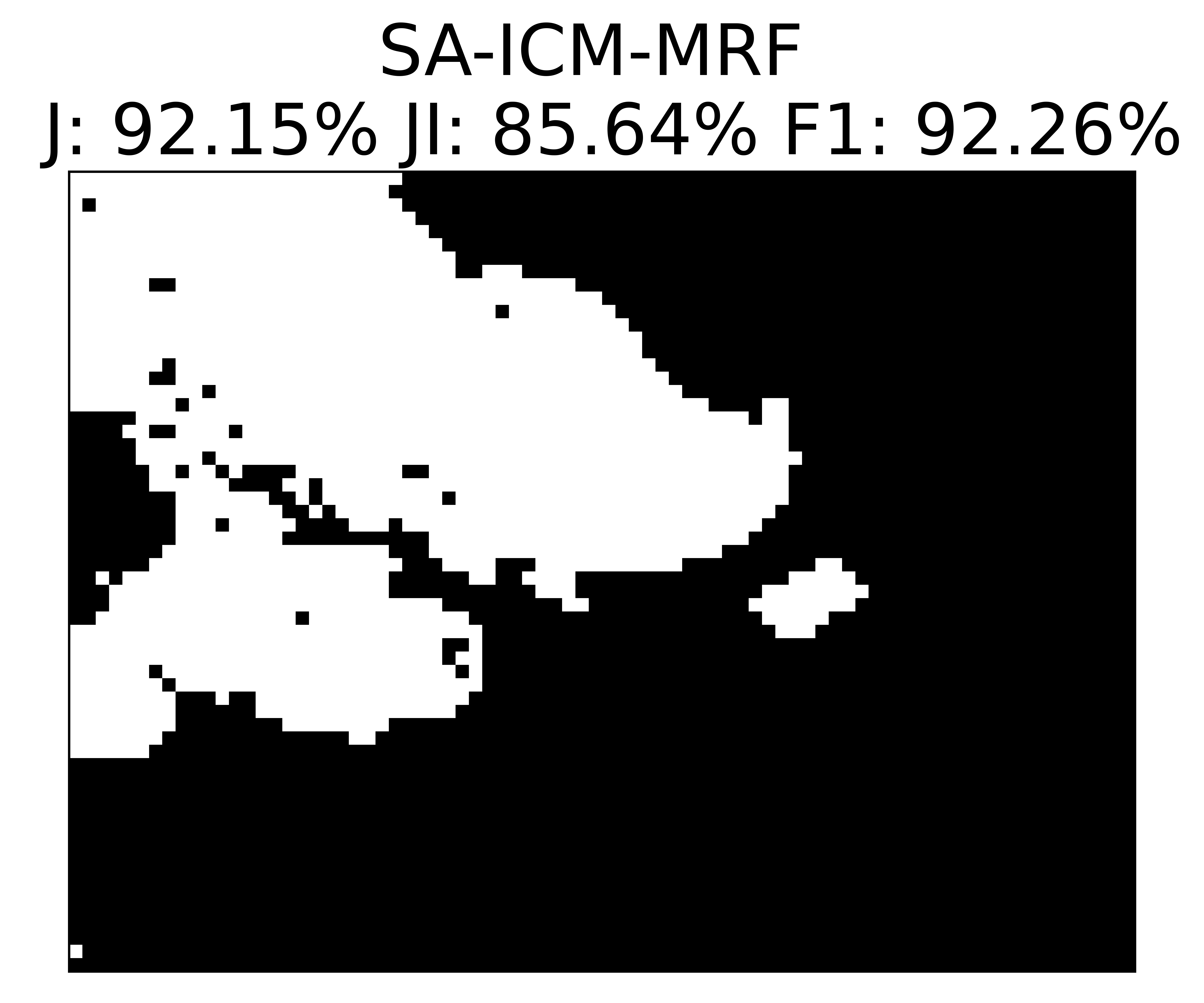}
        \includegraphics[scale = .35, trim = 4 4 4 4, clip]{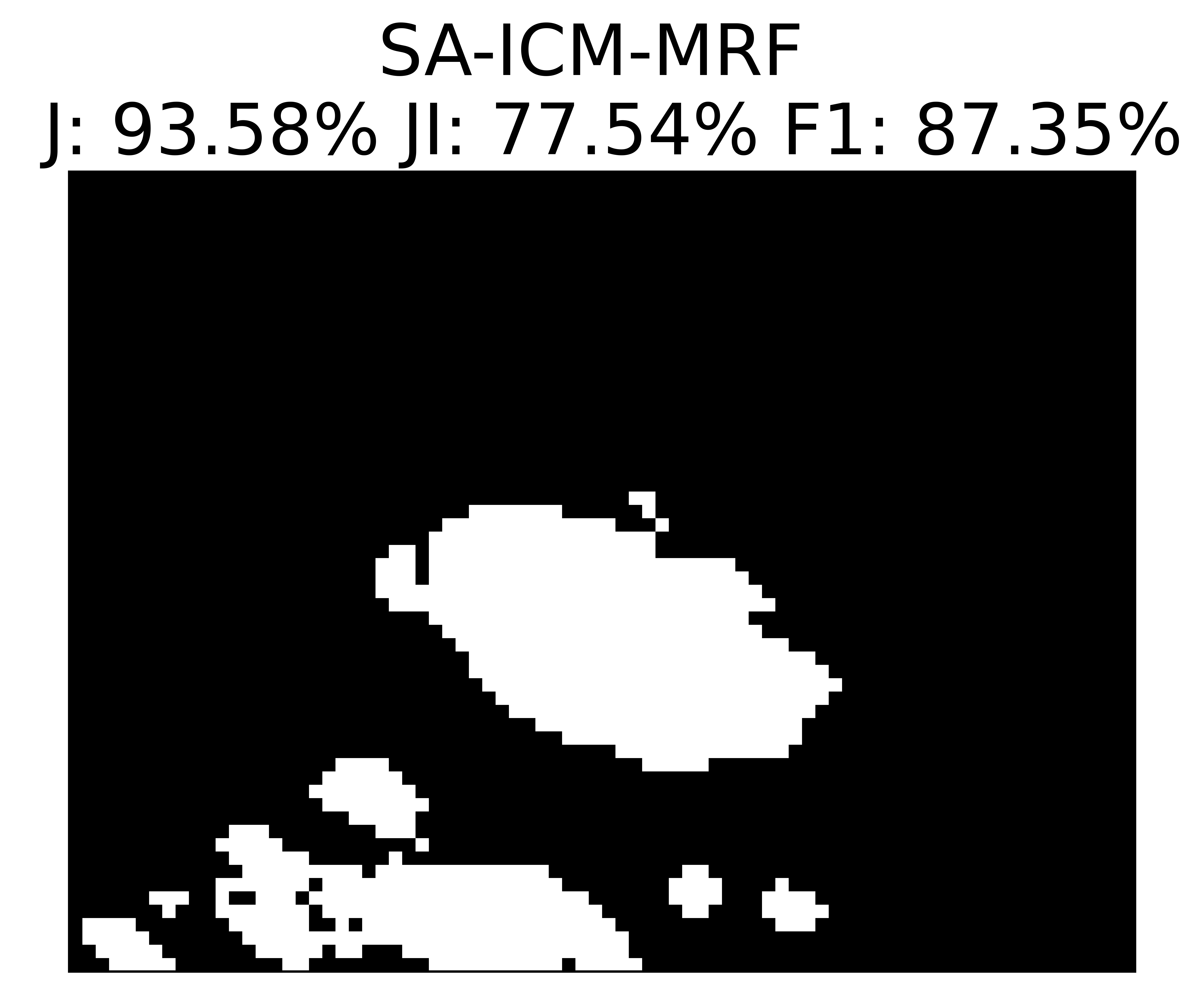}
    \end{subfigure}
    \caption{This figure shows three images from the testing set in the columns. The rows are the segmentation performed by the models. The j-statistic (J), Jaccard Index (JI) and F1 score (F1) achieved by each segmentation model is displayed on the top of the testing images. The higher j-statistic was achieved by ICM-MRF in the first image and by GMM in the second and third image.}
    \label{fig:mrf_test}
\end{figure}

The set of hyperparameter is different for each model, k-means clustering does not have hyperparameter, but the GMM has the covariance matrix regularization term $\varepsilon$ which needs cross-validation in Eq. \eqref{eq:EM_likelihood}. The ICM-MRF and SA-ICM-MRF require the cross-validation of the cliques potential $\beta$ in Eq. \eqref{eq:potential_function}. The ICM algorithm is computationally expensive, so the regularization term of the covariance matrix was set to $\varepsilon = 1$. The cooling hyperparameter of the SA-ICM-MRF was also set to $\alpha = 0.75$. This value of $\alpha$ was found to be an optimal compromise between speed and convergence. In the k-means clustering algorithm, the feature vectors were standardized $\bar{\mathbf{x}}_{i,j} = [ \mathbf{x}_{i,j} - \mathbf{E} (\mathbf{X}) ] / \mathbf{Var} (\mathbf{X})$. The rest of the models neither required normalization nor standardization of the feature vectors, since the covariance matrix is inferred in the learning phase.

ICM-MRF reached the highest j-statistic of 92.55 \% but the average testing time was the highest, at 640.81ms. SA-ICM-MRF achieved a lower j-statistic of 90.06 \% but the implementation of the SA algorithm reduced the average testing time to 136.53ms. The GMM has the best compromise between an average testing time of 4ms and j-statistic of 89.39 \%. The k-means algorithm reached the fastest average testing time of 1ms, but had the lowest j-statistic of 80.73 \%. The preprocessing time of the feature vectors is 0.1 ms for ${\bf x}^1$, 4.7 ms for ${\bf x}^2$, 99.9 ms for ${\bf x}^3$ and 1079 ms for ${\bf x}^4$. When the preprocessing time is considered, the GMM (1083ms) is much slower than the rest of the algorithm. Taking the preprocessing time into account, the optimal models are the ICM-MRF (740.71ms) or the SA-ICM-MRF (236.43ms) depending on the time constraints of the user.

The j-statistic obtained in the testing images is compared to the Jaccard index and F1 score (see Fig. 2). The accuracy metrics are consistent with the exception of the segmentation perfomred by k-means in the second testing image. This inconsistency is due to the challenging segmentation of the selected images, even when the proposed preprocessing is implemented. Nevertheless, the Jaccard index and F1 score achieved by the rest of the models in the first and third image testing images are consistent with j-statistics pointing to the most accurate model.

The experiments were carried out in the Wheeler high performance computer of the UNM-CARC, which uses a SGI AltixXE Xeon X5550 at 2.67GHz with 6 GB of RAM memory per core, 8 cores per node.

\section{Conclusion}

This investigation aims to find an optimal learning algorithm for real-time segmentation of clouds in thermal images acquired using an infrared sky-imaging system. The thermal images were preprocessed to extract the most informative features for segmentation. Preprocessing removes the scattering effect produced by debris in the outdoor germanium window of the camera, and the direct and scattered irradiance produced by the Sun and the atmosphere.

The performance of the classification models increases when the extracted features were preprocessed and information from neighboring pixels was included in the feature vectors. Further research could investigate the performances of supervised Bayesian methods for cloud segmentation, and compare the classification performances between generative and discriminative models.

\section*{Acknowledgment}

Partially supported by NSF EPSCoR grant number OIA-1757207 and the King Felipe VI Endowed Chair of the University of New Mexico. Authors would like to thank the UNM Center for Advanced Research Computing (CARC) for providing the high performance computing and large-scale storage resources used in this work.

\bibliographystyle{unsrt}  
\bibliography{mybibfile}

\begin{thebibliography}{10}

\bibitem{LAPPALAINEN2017}
Kari Lappalainen and Seppo Valkealahti.
\newblock Output power variation of different pv array configurations during
  irradiance transitions caused by moving clouds.
\newblock {\em Applied Energy}, 190:902 -- 910, 2017.

\bibitem{URQUHART2013}
Bryan Urquhart, Mohamed Ghonima, Dung~(Andu) Nguyen, Ben Kurtz, Chi~Wai Chow,
  and Jan Kleissl.
\newblock Chapter 9 - sky-imaging systems for short-term forecasting.
\newblock In Jan Kleissl, editor, {\em Solar Energy Forecasting and Resource
  Assessment}, pages 195--232. Academic Press, Boston, 2013.

\bibitem{GARCIA2018}
O.~García-Hinde, G.~Terrén-Serrano, M.Á. Hombrados-Herrera,
  V.~Gómez-Verdejo, S.~Jiménez-Fernández, C.~Casanova-Mateo, J.~Sanz-Justo,
  M.~Martínez-Ramón, and S.~Salcedo-Sanz.
\newblock Evaluation of dimensionality reduction methods applied to numerical
  weather models for solar radiation forecasting.
\newblock {\em Engineering Applications of Artificial Intelligence}, 69:157 --
  167, 2018.

\bibitem{SOBRI2018}
Sobrina Sobri, Sam Koohi-Kamali, and Nasrudin~Abd. Rahim.
\newblock Solar photovoltaic generation forecasting methods: A review.
\newblock {\em Energy Conversion and Management}, 156:459--497, 2018.

\bibitem{KONG2020}
Weicong Kong, Youwei Jia, Zhao~Yang Dong, Ke~Meng, and Songjian Chai.
\newblock Hybrid approaches based on deep whole-sky-image learning to
  photovoltaic generation forecasting.
\newblock {\em Applied Energy}, 280:115875, 2020.

\bibitem{MAMMOLI2019}
Andrea Mammoli, Guillermo Terren-Serrano, Anthony Menicucci, Thomas~P Caudell,
  and Manel Mart{\'\i}nez-Ram{\'o}n.
\newblock An experimental method to merge far-field images from multiple
  longwave infrared sensors for short-term solar forecasting.
\newblock {\em Solar Energy}, 187:254--260, 2019.

\bibitem{Escrig2013}
H.~Escrig, Francisco Batlles, Joaquín Alonso-Montesinos, F.M. Baena, Juan
  Bosch, I.~Salbidegoitia, and Juan Burgaleta.
\newblock Cloud detection, classification and motion estimation using
  geostationary satellite imagery for cloud cover forecast.
\newblock {\em Energy}, 55, 06 2013.

\bibitem{SHAW2013}
Joseph~A Shaw and Paul~W Nugent.
\newblock Physics principles in radiometric infrared imaging of clouds in the
  atmosphere.
\newblock {\em European Journal of Physics}, 34(6):S111--S121, oct 2013.

\bibitem{THURAIRAJAH2007}
B.~{Thurairajah} and J.~A. {Shaw}.
\newblock Cloud statistics measured with the infrared cloud imager.
\newblock {\em IEEE Transactions on Geoscience and Remote Sensing},
  43(9):2000--2007, Sep. 2005.

\bibitem{NUGENT2009}
Paul~W. Nugent, Joseph~A. Shaw, and Sabino Piazzolla.
\newblock Infrared cloud imaging in support of earth-space optical
  communication.
\newblock {\em Opt. Express}, 17(10):7862--7872, May 2009.

\bibitem{BERNECKER2014}
David Bernecker, Christian Riess, Elli Angelopoulou, and Joachim Hornegger.
\newblock Continuous short-term irradiance forecasts using sky images.
\newblock {\em Solar Energy}, 110:303--315, 2014.

\bibitem{Shi2017}
Cunzhao Shi, Yu~Wang, Chunheng Wang, and Baihua Xiao.
\newblock Ground-based cloud detection using graph model built upon
  superpixels.
\newblock {\em IEEE Geoscience and Remote Sensing Letters}, 14(5):719--723,
  2017.

\bibitem{shawe2004}
John Shawe-Taylor, Nello Cristianini, et~al.
\newblock {\em Kernel methods for pattern analysis}.
\newblock Cambridge university press, 2004.

\bibitem{Taravat2015}
Alireza Taravat, F.~Del~Frate, Cristina Cornaro, and Stefania Vergari.
\newblock Neural networks and support vector machine algorithms for automatic
  cloud classification of whole-sky ground-based images.
\newblock {\em IEEE Geoscience and Remote Sensing Letters}, 12, 02 2015.

\bibitem{Deng2019}
C.~{Deng}, Z.~{Li}, W.~{Wang}, S.~{Wang}, L.~{Tang}, and A.~C. {Bovik}.
\newblock Cloud detection in satellite images based on natural scene statistics
  and gabor features.
\newblock {\em IEEE Geoscience and Remote Sensing Letters}, 16(4):608--612,
  April 2019.

\bibitem{Dronner2018}
Johannes Drönner, Nikolaus Korfhage, Sebastian Egli, Markus Mühling, Boris
  Thies, Jörg Bendix, Bernd Freisleben, and Bernhard Seeger.
\newblock Fast cloud segmentation using convolutional neural networks.
\newblock {\em Remote Sensing}, 10:1782, 11 2018.

\bibitem{Ester1996}
Martin Ester, Hans-Peter Kriegel, J\"{o}rg Sander, and Xiaowei Xu.
\newblock A density-based algorithm for discovering clusters in large spatial
  databases with noise.
\newblock In {\em Proceedings of the Second International Conference on
  Knowledge Discovery and Data Mining}, KDD'96, page 226–231. AAAI Press,
  1996.

\bibitem{mcinnes2017}
Leland McInnes, John Healy, and Steve Astels.
\newblock hdbscan: Hierarchical density based clustering.
\newblock {\em Journal of Open Source Software}, 2(11):205, 2017.

\bibitem{HAR2005}
Sariel Har-Peled and Bardia Sadri.
\newblock How fast is the <i>k</i>-means method?
\newblock In {\em Proceedings of the Sixteenth Annual ACM-SIAM Symposium on
  Discrete Algorithms}, SODA '05, page 877–885, USA, 2005. Society for
  Industrial and Applied Mathematics.

\bibitem{Brigato2021}
Lorenzo Brigato and Luca Iocchi.
\newblock A close look at deep learning with small data.
\newblock In {\em 2020 25th International Conference on Pattern Recognition
  (ICPR)}, pages 2490--2497, 2021.

\bibitem{MURPHY2012}
Kevin~P Murphy.
\newblock {\em Machine learning: a probabilistic perspective}.
\newblock MIT press, 2012.

\bibitem{HOLTON2013}
James~R. Holton and Gregory~J. Hakim.
\newblock Chapter 2 - basic conservation laws.
\newblock In {\em An Introduction to Dynamic Meteorology (Fifth Edition)},
  pages 31 -- 66. Academic Press, 5 edition, 2013.

\bibitem{TERREN2020a}
Guillermo Terrén-Serrano and Manel Martínez-Ramón.
\newblock Data acquisition and image processing for solar irradiance forecast,
  2020.

\bibitem{Hummel1981}
JR~Hummel and WR~Kuhn.
\newblock Comparison of radiative-convective models with constant and
  pressure-dependent lapse rates.
\newblock {\em Tellus}, 33(3):254--261, 1981.

\bibitem{Lucas1981}
B.~D. Lucas and T.~Kanade.
\newblock An iterative image registration technique with an application to
  stereo vision.
\newblock In {\em IJCAI}, 1981.

\bibitem{STAN2001}
Stan~Z. Li.
\newblock {\em {M}arkov Random Field Modeling in Image Analysis}.
\newblock Springer-Verlag, Berlin, Heidelberg, 2001.

\bibitem{HAMMERSLEY1971}
J.~M. Hammersley and P.~Clifford.
\newblock {M}arkov fields on finite graphs and lattices.
\newblock Unpublished, 1971.

\bibitem{BESAG1986}
Julian Besag.
\newblock On the statistical analysis of dirty pictures.
\newblock {\em Journal of the Royal Statistical Society B}, 48(3):48--259,
  1986.

\bibitem{KIRK1983}
S.~Kirkpatrick, C.~D. Gelatt, and M.~P. Vecchi.
\newblock Optimization by simulated annealing.
\newblock {\em Science}, 220(4598):671--680, 1983.

\bibitem{KATO2001}
Zoltan Kato and Ting-Chuen Pong.
\newblock A {M}arkov random field image segmentation model using combined color
  and texture features.
\newblock In W{\l}adys{\l}aw Skarbek, editor, {\em Computer Analysis of Images
  and Patterns}, pages 547--554. Springer Berlin Heidelberg, 2001.

\bibitem{YOUDEN1950}
W.~J. Youden.
\newblock Index for rating diagnostic tests.
\newblock {\em Cancer}, 3(1):32--35, 1950.

\bibitem{FAWCETT2006}
Tom Fawcett.
\newblock An introduction to {ROC} analysis.
\newblock {\em Pattern Recognition Letters}, 27(8):861--874, June 2006.

\bibitem{TERREN2020d}
Guillermo Terrén-Serrano, Adnan Bashir, Trilce Estrada, and Manel
  Martínez-Ramón.
\newblock Girasol, a sky imaging and global solar irradiance dataset.
\newblock {\em Data in Brief}, page 106914, 2021.

\end{thebibliography}

\end{document}